\documentclass[reprint,amsmath,amssymb,aps,showkeys,onecolumn]{revtex4-2}
\usepackage{graphicx}
\usepackage{dcolumn}
\usepackage{bm}
\usepackage{ulem}
\usepackage{amsmath}
\usepackage{amssymb}
\usepackage{mathrsfs}
\usepackage{helvet}
\usepackage{courier}
\usepackage{kbordermatrix}
\usepackage{fancybox,ascmac}
\usepackage{multirow,bigdelim}
\usepackage[legacycolonsymbols]{mathtools}

\usepackage{color}
\usepackage{mathrsfs}

\begin{document}


\title{Magnon Confinement on the Two-Dimensional Penrose Lattice:\\
       Perpendicular-Space Analysis of the Dynamic Structure Factor}

\author{Shoji Yamamoto${}^{*}$ and Takashi Inoue}
\affiliation{Department of Physics, Hokkaido University, Sapporo 060-0810, Japan} 

\begin{abstract}
   Employing the spin-wave formalism within and beyond the harmonic-oscillator approximation,
we study the dynamic structure factors of spin-$\frac{1}{2}$ nearest-neighbor quantum Heisenberg
antiferromagnets on two-dimensional quasiperiodic lattices with particular emphasis on a magnetic
analog to the well-known confined states of a hopping Hamiltonian for independent electrons on
a two-dimensional Penrose lattice.
We present comprehensive calculations on the $\mathbf{C}_{5\mathrm{v}}$ Penrose tiling
in comparison with the $\mathbf{C}_{8\mathrm{v}}$ Ammann-Beenker tiling, revealing their
decagonal and octagonal antiferromagnetic microstructures.
Their dynamic spin structure factors both exhibit linear soft modes emergent at magnetic Bragg
wavevectors and have nearly or fairly flat scattering bands, signifying magnetic excitations
localized in some way, at several different energies in a self-similar manner.
In particular, the lowest-lying highly flat mode is distinctive of the Penrose lattice,
which is mediated by its unique antiferromagnons confined within tricoordinated sites only,
unlike their itinerant electron counterparts involving pentacoordinated as well as tricoordinated
sites.
Bringing harmonic antiferromagnons into higher-order quantum interaction splits
the lowest-lying nearly flat scattering band in two, each mediated by further confined
antiferromagnons, which is fully demonstrated and throughly visualized in the perpendicular
as well as real spaces.
We disclose \textit{superconfined antiferromagnons} on the two-dimensional Penrose lattice.
\end{abstract}

\maketitle


\section{Introduction}

\begin{figure}
\centering
\includegraphics[width=\linewidth]{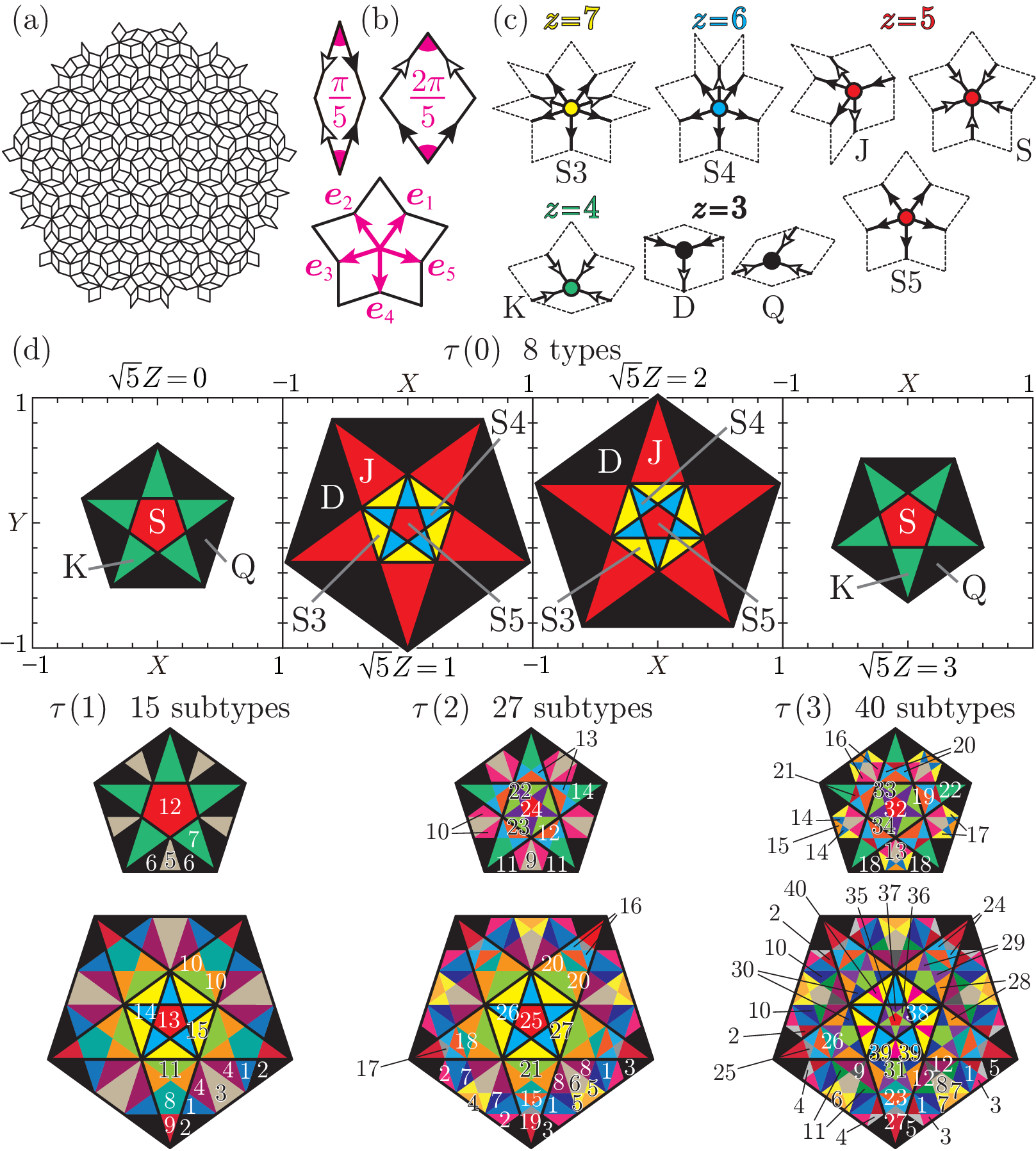}
\caption{%
         (a) A two-dimensional Penrose lattice of $L=526$ with fivefold rotational symmetry.
         (b) The Penrose tiling constitutes of two rhombi with angles $\frac{\pi}{5}$ and
             $\frac{2\pi}{5}$, respectively, whose edges each are marked with an arrow
             so as to define the matching rules.
             The canonical basis vectors of a five-dimensional hypercubic lattice convert into
             five vectors which we shall denote by
             $\bm{e}_{1}$, $\bm{e}_{2}$, $\bm{e}_{3}$, $\bm{e}_{4}$, and $\bm{e}_{5}$.
             We can choose any four among them as the primitive translation vectors for
             the two-dimensional Penrose tiling.
             Note that $\sum_{l=1}^5\bm{e}_l=\bm{0}$.
         (c) Eight types of local environments on the Penrose lattice \cite{dB39, dB53}
             with their coordination numbers $z$ ranging from $3$ to $7$.
         (d) The perpendicular space of the Penrose lattice consists of a three-dimensional
             stack of four pentagons lying at $Z=0,1/\sqrt{5},2/\sqrt{5},3/\sqrt{5}$,
             which are divided into several domains colored differently, each containing the same
             species of vertices only and therefore being labeled any of the eight species
             $\tau(0)=\mathrm{D}$ to S3.
             Such labeled domains in the perpendicular space are subdivided according to
             their surrounding environments in the physical space.
             Suppose we define the distance-$R$ surrounding environment as all the vertices within
             the distance $R$.
             Then the $8$ types of domains at $R=0$ divide into $15$, $27$, and $40$ types of
             subdomains with $R$ being set to $1$, $2$, and $3$, respectively.
             Refer to Tables \ref{T:vertexR1}--\ref{T:vertexR3} for the way we number
             the local environments.}
\label{F:Penrose}
\end{figure}

   In 1974, Penrose \cite{P266} discovered an extreme manner of non-periodic tiling
in two dimensions [Figure \ref{F:Penrose}(a)].
What we call the Penrose lattice consists of two rhombi with angles of 
$\frac{\pi}{5}$ and $\frac{2\pi}{5}$ laid out with a matching rule but without any translational
symmetry [Figure \ref{F:Penrose}(b)].
The Penrose lattice has a crystallographically forbidden five-fold rotational symmetry and 
a self-similar structure characterized by the golden ratio $\frac{1+\sqrt{5}}{2}$ \cite{G110}.
The diffraction pattern of the Penrose lattice, however, consists of sharp $\delta$-function
peaks \cite{M609}.
Hence it follows that the Penrose lattice has a long-range positional order, which we nowadays
refer to as ``quasiperiodicity".
Besides the Penrose tile pattern, numerous ways of quasiperiodic tiling have been proposed to date,
with eight-fold \cite{B82wsk04} and twelve-fold \cite{NL405} rotational symmetries for instance.
Such unconventional geometries fascinate physicists as well as mathematicians.
Several authors took their early interest in tight-binding models \cite{C2915,O2184,T1420,T8879}
on the Penrose lattice.
When electrons hop on the vertices of each rhombus, they yield a macroscopically
degenerate eigenlevel of zero energy in the thermodynamic limit \cite{K2740,A1621,Z3377}.
While wavefunctions of these unique states are all strictly confined in a finite region,
they take on a critical character, i.e., they are neither localized nor extended in
the conventional sense to have no absolute length scale.
The density of states consists of spiky bands and, in particular, the zero-energy confined states
constitute a $\delta$-function peak, being isolated from all the rest by a gap of about a tenth of
the hopping energy $t$.
All these findings are consequent from the self-similarity of the Penrose lattice.
\begin{figure}
\centering
\includegraphics[width=0.75\linewidth]{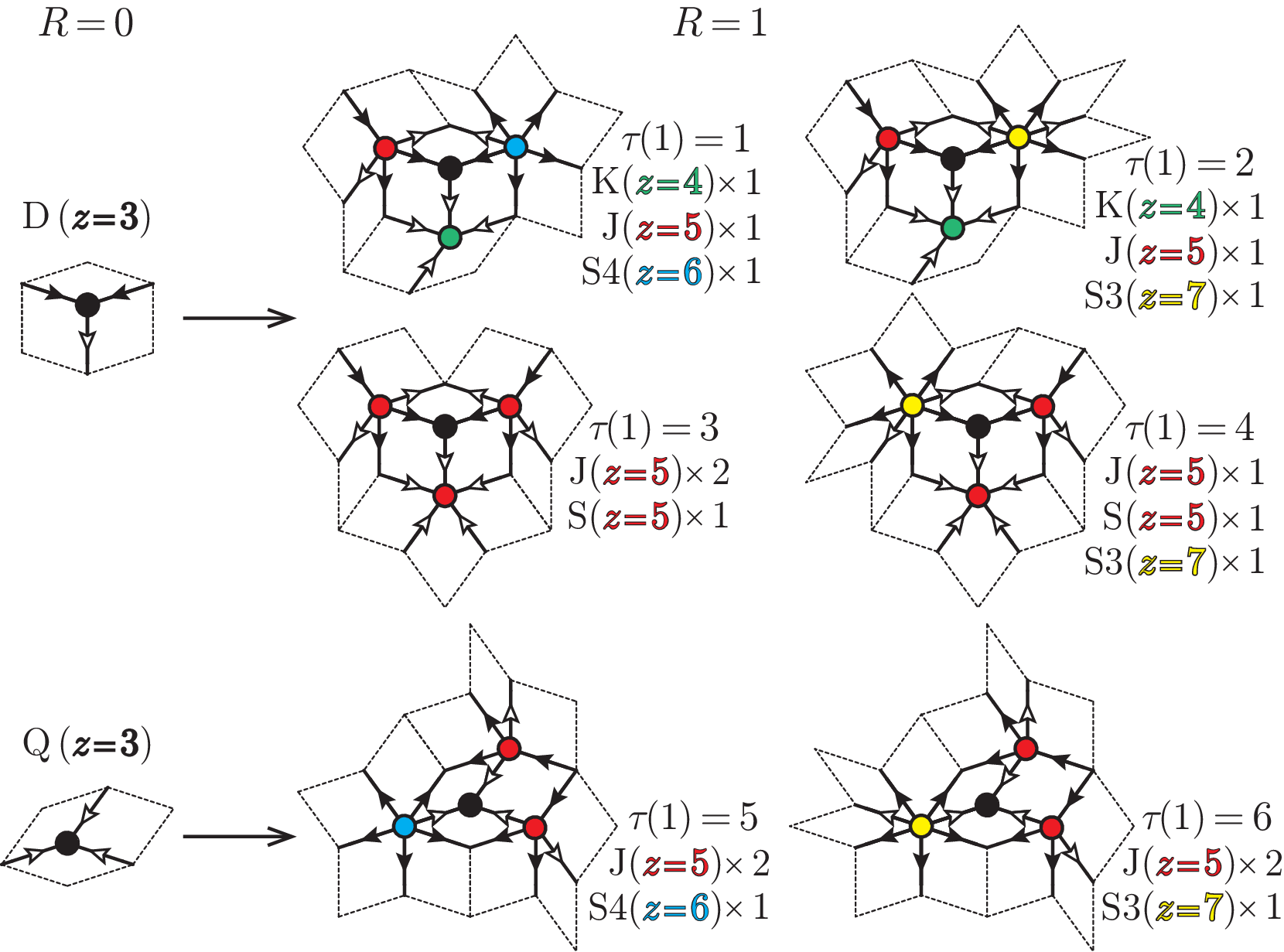}
\caption{%
         Subclassifications of tricoordinated vertices on the two-dimensional Penrose lattice.
         When we define their surrounding environments as their nearest neighbors, i.e.,
         surrounding vertices at a distance of $R=1$, the species D and Q quadrifurcate and
         bifurcate, respectively.
         Refer to Tables \ref{T:vertexR1}--\ref{T:vertexR3} for further details.}
\label{F:local_exDandQ}
\end{figure}

   The synthesis of an aluminum-manganese alloy of icosahedral point group symmetry \cite{S1951}
surprised us in a different context.
Indeed its novel rotational symmetry inconsistent with any translational symmetry is
suggestive of quasiperiodic tiling, but it was left out of solids,
before arguing whether this should also come under crystals,
due to its lack of thermodynamic stability.
However, thermodynamically stable quasicrystals \cite{TL1505, TL1587} were synthesized a few years
later and it did not take much time until the concept and definition of ``quasicrystals"
\cite{L2477, L596} were established.
The birth of quasicrystals caused a revolution in materials science, sparking a renewed
interest in the Penrose lattice.
Correlated electrons on the two-dimensional Penrose lattice have been investigated in terms of
pure \cite{K214402} and extended \cite{S054202} Hubbard models.
Without any Coulomb interaction, the ground-state charge distribution is uniform
at half band filling.
Once the system deviates from its half filling, the charge distribution is no longer uniform,
which is the consequence of various constituent coordination numbers $z$, ranging
from $3$ to $7$.
Coulomb interactions lift the degeneracy of the confined states and induce a magnetic or
charge order.
At half filling, an infinitesimal on-site Coulomb repulsion $U$ induces nonzero local staggered
magnetizations because of the macroscopically degenerate zero-energy states.
With increasing $U$, the confined states come to constitute a band of width in proportion to $U$
provided $U\ll t$.
The intersite Coulomb repulsion $V$ breaks the electron-hole symmetry and serves to induce charge
inhomogeneity rather than any magnetic ordering.
If we dope the system with holes or electrons, it becomes metallic with its confined states being
totally vacant or fully occupied, and then its ground state is charge ordered in a nonuniform and
self-similar manner.

   In the strong-correlation limit, the half-filled single-band Hubbard model on an arbitrary
lattice behaves as a spin-$\frac{1}{2}$ Heisenberg antiferromagnet.
Within the naivest spin-wave (SW) formalism, Wessel and Milat \cite{W104427} calculated
the dynamic structure factor, accessible to inelastic neutron scattering, of the nearest-neighbor
antiferromagnetic spin-$\frac{1}{2}$ Heisenberg model on the Ammann-Beenker octagonal tiling
\cite{B82wsk04}, which is another prominent example of a quasiperiodic lattice in two dimensions.
They found linear soft modes near the magnetic Bragg peaks at low frequencies, 
self-similar structures at intermediate frequencies, and flat bands at high frequencies.
These findings each are intriguing in themselves and more intriguing is their coexistence
that has never happened in any conventional periodic system.
The bipartiteness of two-dimensional quasiperiodic tilings such as the Penrose and Ammann-Beenker
lattices allows quantum Monte Carlo simulations of antiferromagnetically exchange-coupled spins
residing on them \cite{J212407,W177205}.
As in the case of the square-lattice antiferromagnet, they have N\'eel-ordered ground states,
whose local staggered magnetizations, however, vary site by site.
The spatially averaged staggered magnetizations on the Penrose \cite{J212407} and Ammann-Beenker
\cite{W177205} lattices are both larger than that on the square lattice \cite{S11678}, though
they can be marked by the same coordination number $z=4$ on an average.
This reads as the spatial inhomogeneity serving to reduce quantum fluctuations.
The linear SW (LSW) theory well reproduces their bulk properties in general.
Its estimates of the ground-state energy \cite{W104427,J212407} are fairly accurate with
an error less than $2$ percent.
The LSW theory securely guesses the ground-state staggered magnetizations as well when averaged
spatially, but its findings for their local values each are no longer reliable quantitatively
\cite{W104427,S104427}.
We must go beyond the harmonic-oscillator approximation in an attempt to reveal the very nature of
quasiperiodic quantum magnets.
Although the Penrose and Ammann-Beenker tilings have coordination numbers $z$ ranging from
$3$ to $7$ and $3$ to $8$, respectively, the former has more local environments
[Figure \ref{F:Penrose}(c)] and more complicated transformation rules than the latter.
There may be differences as well as similarities between them, as was demonstrated in
inelastic-light-scattering studies \cite{I2000118,I053701}.
Thus motivated, we investigate static and dynamic spin structure factors for the nearest-neighbor
antiferromagnetic spin-$\frac{1}{2}$ Heisenberg model on the Penrose lattice, in comparison with
those on the Ammann-Beenker lattice, in terms of the SW language within and beyond
the harmonic-oscillator approximation.

   $3d$-transition-metal-based icosahedral quasicrystals such as
Zn-Mg-Ho \cite{S476} and Zn-Mg-Tb \cite{S054417} have extensively been studied through
the use of neutron scattering \cite{S476,S054417,S014437}, because their atomic structure
is relatively well known and their sizable $4f$ magnetic moments are well localized and
mostly isotropic.
Recent technical progress of manipulating optical lattices \cite{G3363,V110404,S200604}
more and more motivate our theoretical investigations.
Two-dimensional quasiperiodic optical potentials of five- \cite{SP053607} or eight-fold
\cite{J66003,J149} rotational invariance were designed in terms of standing-wave lasers
and consequent quasiperiodic long-range orders indeed manifested in sharp diffraction peaks
\cite{SP053607,V110404}.
By trapping ultracold bosonic or fermionic atoms in optical potentials and controlling
the intensity, frequency, and polarization of the laser beam, we can change the on-site
interaction and tunneling energy to obtain various effective spin Hamiltonians \cite{D090402}.

\section{Quasiperiodic Tilings in Two Dimensions}
   Figure \ref{F:Penrose}(a) is a Penrose lattice in two dimensions, which can be generated
from minimal seeds [Figure \ref{F:Penrose}(b)] by matching up their edges of the same arrow and
repeating the inflation-deflation operation \cite{K214402} with a magnification ratio of
the golden number $\frac{1+\sqrt{5}}{2}$ to unity.
Such operations yield the self-similarity of the Penrose lattice.
Every two vertices on the Penrose lattice are connected by a linear combination of four independent
primitive lattice vectors [Figure \ref{F:Penrose}(b)] with a unique set of four integral
coefficients.
While the actual physical dimension $d$ of the Penrose lattice equals $2$,
its indexing dimension or simply rank $D$ amounts to $4$ \cite{L313}.
We can generally obtain a physically $d$-dimensional quasiperiodic tiling from
a higher-than-$d$-dimensional periodic lattice through an algebraic approach \cite{dB39, dB53}.
The Penrose lattice is obtained as a projection of a five-dimensional hypercubic lattice onto
a two-dimensional physical space,
\begin{align}
   \left[
   \begin{array}{c}
    x \\
    y \\
    X \\
    Y \\
    Z
   \end{array}
   \right]
  =
   \sqrt{\frac{2}{5}}
   \left[
   \begin{array}{ccccc}
    \cos\phi_{0} & \cos\phi_{1} & \cos\phi_{2} & \cos\phi_{3} & \cos\phi_{4} \\
    \sin\phi_{0} & \sin\phi_{1} & \sin\phi_{2} & \sin\phi_{3} & \sin\phi_{4} \\
    \cos\phi_{0} & \cos\phi_{2} & \cos\phi_{4} & \cos\phi_{6} & \cos\phi_{8} \\
    \sin\phi_{0} & \sin\phi_{2} & \sin\phi_{4} & \sin\phi_{6} & \sin\phi_{8} \\
    \frac{1}{\sqrt{2}} & \frac{1}{\sqrt{2}} & \frac{1}{\sqrt{2}} & \frac{1}{\sqrt{2}} & \frac{1}{\sqrt{2}}
   \end{array}
   \right]
   \left[
   \begin{array}{c}
    m_{1} \\
    m_{2} \\
    m_{3} \\
    m_{4} \\ 
    m_{5}
   \end{array}
   \right]
\ \ \ 
   \left(
    \phi_{n}
   \equiv
    \frac{2\pi}{5}n+\frac{\pi}{10},\ 
   m_n\in\mathbb{Z}
   \right),
   \label{E:projectionPenrose}
\end{align}
where $(x,y)$ and $(X,Y,Z)$ are the coordinates in the two-dimensional physical and
three-dimensional perpendicular spaces, respectively, and
the prefactor $\sqrt{2/5}$ plays a lattice constant in the former.
The physical lattice and therefore its perpendicular space each have an irrational gradient to
the higher-dimensional periodic lattice and therefore points in the former and latter are brought
into one-to-one correspondence.
\begin{table}
\caption{%
         The way of numbering local environments of $R=1$ for the Penrose lattice.
         All the vertices are classified into $15$ subtypes and labeled $\tau(1)=1$ to $15$,
         according as how many vertices of the types D to S3 appear at distances $R=0$ and $1$,
         when we truncate their surrounding vertices at a distance of $R=1$.
         In an attempt to characterize each environment, we sum up the coordination numbers of
         nearest-neighbor ($R=1$) sites at $\bm{r}_{l'}$ to its core ($R=0$) site at $\bm{r}_{l}$.}
\begin{tabular*}{\linewidth}{@{\extracolsep{\fill}} c || c | c | c | c | c | c | c | c || c }
\hline\hline
$\tau(1)$
  & D($z=3$) & Q($z=3$) & K($z=4$) & J($z=5$) & S($z=5$) & S5($z=5$) & S4($z=6$) & S3($z=7$) &
$\sum_{l'}z_{l'}(1)$
\\
\hline\hline
1 & (1;0) & (0;0) & (0;1) & (0;1) & (0;0) & (0;0) & (0;1) & (0;0) & 15
\\
\hline
2 & (1;0) & (0;0) & (0;1) & (0;1) & (0;0) & (0;0) & (0;0) & (0;1) & 16
\\
\hline
3 & (1;0) & (0;0) & (0;0) & (0;2) & (0;1) & (0;0) & (0;0) & (0;0) & 15
\\
\hline
4 & (1;0) & (0;0) & (0;0) & (0;1) & (0;1) & (0;0) & (0;0) & (0;1) & 17
\\
\hline
5 & (0;0) & (1;0) & (0;0) & (0;2) & (0;0) & (0;0) & (0;1) & (0;0) & 16
\\
\hline
6 & (0;0) & (1;0) & (0;0) & (0;2) & (0;0) & (0;0) & (0;0) & (0;1) & 17
\\
\hline
7 & (0;2) & (0;0) & (1;0) & (0;2) & (0;0) & (0;0) & (0;0) & (0;0) & 16
\\
\hline
8 & (0;2) & (0;2) & (0;0) & (1;0) & (0;0) & (0;1) & (0;0) & (0;0) & 17
\\
\hline
9 & (0;2) & (0;2) & (0;0) & (1;0) & (0;0) & (0;0) & (0;1) & (0;0) & 18
\\
\hline
10 & (0;2) & (0;1) & (0;1) & (1;0) & (0;0) & (0;0) & (0;1) & (0;0) & 19
\\
\hline
11 & (0;2) & (0;0) & (0;2) & (1;0) & (0;0) & (0;0) & (0;0) & (0;1) & 21
\\
\hline
12 & (0;5) & (0;0) & (0;0) & (0;0) & (1;0) & (0;0) & (0;0) & (0;0) & 15
\\
\hline
13 & (0;0) & (0;0) & (0;0) & (0;5) & (0;0) & (1;0) & (0;0) & (0;0) & 25
\\
\hline
14 & (0;2) & (0;1) & (0;0) & (0;3) & (0;0) & (0;0) & (1;0) & (0;0) & 24
\\
\hline
15 & (0;4) & (0;2) & (0;0) & (0;1) & (0;0) & (0;0) & (0;0) & (1;0) & 23
\\
\hline\hline
\end{tabular*}
\label{T:vertexR1}
\end{table}
\begin{table}
\caption{%
         The way of numbering local environments of $R=2$ for the Penrose lattice.
         All the vertices are classified into $27$ subtypes and labeled $\tau(2)=1$ to $27$,
         according as how many vertices of the types D to S3 appear at distances $R=0$, $1$,
         and $2$, when we truncate their surrounding vertices at a distance of $R=2$.
         In an attempt to characterize each environment, we sum up the coordination numbers of
         nearest-neighbor ($R=1$) sites at $\bm{r}_{l'}$ to its core ($R=0$) site at $\bm{r}_{l}$.}
\begin{tabular*}{\linewidth}{@{\extracolsep{\fill}} c || c | c | c | c | c | c | c | c || c }
\hline\hline
$\tau(2)$
  & D($z=3$) & Q($z=3$) & K($z=4$) & J($z=5$) & S($z=5$) & S5($z=5$) & S4($z=6$) & S3($z=7$) &
$\sum_{l'}z_{l'}(1)$     
\\
\hline\hline
1 &(1;0,2)&(0;0,2)&(0;1,0)&(0;1,4)&(0;0,0)&(0;0,1)&(0;1,0)&(0;0,0) & 15
\\
\hline
2 &(1;0,4)&(0;0,3)&(0;1,0)&(0;1,2)&(0;0,0)&(0;0,1)&(0;0,0)&(0;1,0) & 16
\\
\hline
3 &(1;0,4)&(0;0,3)&(0;1,0)&(0;1,2)&(0;0,0)&(0;0,0)&(0;0,1)&(0;1,0) & 16
\\
\hline
4 &(1;0,4)&(0;0,2)&(0;0,1)&(0;2,0)&(0;1,0)&(0;0,0)&(0;0,2)&(0;0,0) & 15
\\
\hline
5 &(1;0,4)&(0;0,1)&(0;0,2)&(0;2,0)&(0;1,0)&(0;0,0)&(0;0,1)&(0;0,1) & 15
\\
\hline
6 &(1;0,4)&(0;0,0)&(0;0,3)&(0;2,0)&(0;1,0)&(0;0,0)&(0;0,0)&(0;0,2) & 15
\\
\hline
7 &(1;0,6)&(0;0,3)&(0;0,0)&(0;1,1)&(0;1,0)&(0;0,1)&(0;0,0)&(0;1,0) & 17
\\
\hline
8 &(1;0,6)&(0;0,2)&(0;0,1)&(0;1,1)&(0;1,0)&(0;0,0)&(0;0,1)&(0;1,0) & 17
\\
\hline
9 &(0;0,4)&(1;0,2)&(0;0,0)&(0;2,3)&(0;0,0)&(0;0,1)&(0;1,0)&(0;0,0) & 16
\\
\hline
10 &(0;0,6)&(1;0,3)&(0;0,0)&(0;2,1)&(0;0,0)&(0;0,1)&(0;0,0)&(0;1,0) & 17
\\
\hline
11 &(0;0,6)&(1;0,2)&(0;0,1)&(0;2,1)&(0;0,0)&(0;0,0)&(0;0,1)&(0;1,0) & 17
\\
\hline
12 &(0;2,3)&(0;0,2)&(1;0,0)&(0;2,1)&(0;0,0)&(0;0,0)&(0;0,2)&(0;0,0) & 16
\\
\hline
13 &(0;2,3)&(0;0,1)&(1;0,1)&(0;2,1)&(0;0,0)&(0;0,0)&(0;0,1)&(0;0,1) & 16
\\
\hline
14 &(0;2,3)&(0;0,0)&(1;0,2)&(0;2,1)&(0;0,0)&(0;0,0)&(0;0,0)&(0;0,2) & 16
\\
\hline
15 &(0;2,0)&(0;2,0)&(0;0,1)&(1;0,4)&(0;0,0)&(0;1,0)&(0;0,2)&(0;0,0) & 17
\\
\hline
16 &(0;2,0)&(0;2,0)&(0;0,1)&(1;0,4)&(0;0,0)&(0;1,0)&(0;0,1)&(0;0,1) & 17
\\
\hline
17 &(0;2,0)&(0;2,0)&(0;0,1)&(1;0,4)&(0;0,0)&(0;1,0)&(0;0,0)&(0;0,2) & 17
\\
\hline
18 &(0;2,0)&(0;2,0)&(0;0,0)&(1;0,4)&(0;0,1)&(0;1,0)&(0;0,0)&(0;0,2) & 17
\\
\hline
19 &(0;2,2)&(0;2,1)&(0;0,1)&(1;0,2)&(0;0,0)&(0;0,0)&(0;1,0)&(0;0,2) & 18
\\
\hline
20 &(0;2,3)&(0;1,1)&(0;1,0)&(1;0,3)&(0;0,1)&(0;0,0)&(0;1,0)&(0;0,1) & 19
\\
\hline
21 &(0;2,6)&(0;0,2)&(0;2,0)&(1;0,2)&(0;0,1)&(0;0,0)&(0;0,0)&(0;1,0) & 21
\\
\hline
22 &(0;5,0)&(0;0,0)&(0;0,0)&(0;0,5)&(1;0,0)&(0;0,0)&(0;0,0)&(0;0,0) & 15
\\
\hline
23 &(0;5,0)&(0;0,0)&(0;0,0)&(0;0,4)&(1;0,0)&(0;0,0)&(0;0,0)&(0;0,1) & 15
\\
\hline
24 &(0;5,0)&(0;0,0)&(0;0,0)&(0;0,3)&(1;0,0)&(0;0,0)&(0;0,0)&(0;0,2) & 15
\\
\hline
25 &(0;0,10)&(0;0,5)&(0;0,0)&(0;5,0)&(0;0,0)&(1;0,0)&(0;0,0)&(0;0,0) & 25
\\
\hline
26 &(0;2,6)&(0;1,2)&(0;0,2)&(0;3,2)&(0;0,0)&(0;0,0)&(1;0,0)&(0;0,0) & 24
\\
\hline
27 &(0;4,2)&(0;2,0)&(0;0,2)&(0;1,4)&(0;0,1)&(0;0,0)&(0;0,0)&(1;0,0) & 23
\\
\hline\hline
\end{tabular*}
\label{T:vertexR2}
\end{table}
\begin{table}
\caption{%
         The way of numbering local environments of $R=3$ for the Penrose lattice.
         All the vertices are classified into $40$ subtypes and labeled $\tau(3)=1$ to $40$,
         according as how many vertices of the types D to S3 appear at distances $R=0$, $1$, $2$,
         and $3$, when we truncate their surrounding vertices at a distance of $R=3$.
         In an attempt to characterize each environment, we sum up the coordination numbers of
         nearest-neighbor ($R=1$) sites at $\bm{r}_{l'}$ to its core ($R=0$) site at $\bm{r}_{l}$.}
\begin{tabular*}{\linewidth}{@{\extracolsep{\fill}} c || c | c | c | c | c | c | c | c || c }
\hline\hline
$\tau(3)$
  & D($z=3$) & Q($z=3$) & K($z=4$) & J($z=5$) & S($z=5$) & S5($z=5$) & S4($z=6$) & S3($z=7$) &
$\sum_{l'}z_{l'}(1)$     
\\
\hline\hline
1  &(1;0,2,7)&(0;0,2,3)&(0;1,0,1)&(0;1,4,4)&(0;0,0,0)&(0;0,1,0)&(0;1,0,1)&(0;0,0,0) & 15
\\
\hline
2  &(1;0,2,7)&(0;0,2,2)&(0;1,0,2)&(0;1,4,4)&(0;0,0,0)&(0;0,1,0)&(0;1,0,0)&(0;0,0,1) & 15
\\
\hline
3  &(1;0,4,3)&(0;0,3,1)&(0;1,0,1)&(0;1,2,6)&(0;0,0,1)&(0;0,1,0)&(0;0,0,1)&(0;1,0,0) & 16
\\
\hline
4  &(1;0,4,3)&(0;0,3,0)&(0;1,0,2)&(0;1,2,6)&(0;0,0,1)&(0;0,1,0)&(0;0,0,0)&(0;1,0,1) & 16
\\
\hline
5  &(1;0,4,5)&(0;0,3,1)&(0;1,0,2)&(0;1,2,4)&(0;0,0,1)&(0;0,0,0)&(0;0,1,0)&(0;1,0,1) & 16
\\
\hline
6  &(1;0,4,4)&(0;0,2,2)&(0;0,1,0)&(0;2,0,5)&(0;1,0,0)&(0;0,0,0)&(0;0,2,0)&(0;0,0,2) & 15
\\
\hline
7  &(1;0,4,7)&(0;0,1,3)&(0;0,2,0)&(0;2,0,4)&(0;1,0,0)&(0;0,0,0)&(0;0,1,0)&(0;0,1,1) & 15
\\
\hline
8  &(1;0,4,10)&(0;0,0,4)&(0;0,3,0)&(0;2,0,3)&(0;1,0,0)&(0;0,0,0)&(0;0,0,0)&(0;0,2,0) & 15
\\
\hline
9  &(1;0,4,10)&(0;0,0,4)&(0;0,3,0)&(0;2,0,2)&(0;1,0,0)&(0;0,0,0)&(0;0,0,0)&(0;0,2,1) & 15
\\
\hline
10 &(1;0,6,2)&(0;0,3,0)&(0;0,0,2)&(0;1,1,7)&(0;1,0,0)&(0;0,1,0)&(0;0,0,0)&(0;1,0,1) & 17
\\
\hline
11 &(1;0,6,5)&(0;0,2,1)&(0;0,1,2)&(0;1,1,6)&(0;1,0,0)&(0;0,0,0)&(0;0,1,0)&(0;1,0,0) & 17
\\
\hline
12 &(1;0,6,5)&(0;0,2,1)&(0;0,1,2)&(0;1,1,5)&(0;1,0,0)&(0;0,0,0)&(0;0,1,0)&(0;1,0,1) & 17
\\
\hline
13 &(0;0,4,6)&(1;0,2,2)&(0;0,0,2)&(0;2,3,3)&(0;0,0,0)&(0;0,1,0)&(0;1,0,2)&(0;0,0,0) & 16
\\
\hline
14 &(0;0,4,6)&(1;0,2,2)&(0;0,0,2)&(0;2,3,3)&(0;0,0,0)&(0;0,1,0)&(0;1,0,1)&(0;0,0,1) & 16
\\
\hline
15 &(0;0,4,6)&(1;0,2,2)&(0;0,0,2)&(0;2,3,3)&(0;0,0,0)&(0;0,1,0)&(0;1,0,0)&(0;0,0,2) & 16
\\
\hline
16 &(0;0,6,2)&(1;0,3,0)&(0;0,0,2)&(0;2,1,5)&(0;0,0,1)&(0;0,1,0)&(0;0,0,1)&(0;1,0,1) & 17
\\
\hline
17 &(0;0,6,2)&(1;0,3,0)&(0;0,0,2)&(0;2,1,5)&(0;0,0,1)&(0;0,1,0)&(0;0,0,0)&(0;1,0,2) & 17
\\
\hline
18 &(0;0,6,5)&(1;0,2,1)&(0;0,1,2)&(0;2,1,4)&(0;0,0,1)&(0;0,0,0)&(0;0,1,0)&(0;1,0,1) & 17
\\
\hline
19 &(0;2,3,2)&(0;0,2,2)&(1;0,0,0)&(0;2,1,4)&(0;0,0,1)&(0;0,0,1)&(0;0,2,0)&(0;0,0,2) & 16
\\
\hline
20 &(0;2,3,5)&(0;0,1,3)&(1;0,1,0)&(0;2,1,3)&(0;0,0,1)&(0;0,0,1)&(0;0,1,0)&(0;0,1,1) & 16
\\
\hline
21 &(0;2,3,8)&(0;0,0,4)&(1;0,2,0)&(0;2,1,2)&(0;0,0,1)&(0;0,0,1)&(0;0,0,0)&(0;0,2,0) & 16
\\
\hline
22 &(0;2,3,8)&(0;0,0,4)&(1;0,2,0)&(0;2,1,2)&(0;0,0,1)&(0;0,0,0)&(0;0,0,1)&(0;0,2,0) & 16
\\
\hline
23 &(0;2,0,8)&(0;2,0,3)&(0;0,1,0)&(1;0,4,6)&(0;0,0,0)&(0;1,0,0)&(0;0,2,0)&(0;0,0,0) & 17
\\
\hline
24 &(0;2,0,10)&(0;2,0,4)&(0;0,1,0)&(1;0,4,4)&(0;0,0,0)&(0;1,0,0)&(0;0,1,0)&(0;0,1,0) & 17
\\
\hline
25 &(0;2,0,12)&(0;2,0,5)&(0;0,1,0)&(1;0,4,2)&(0;0,0,0)&(0;1,0,0)&(0;0,0,0)&(0;0,2,0) & 17
\\
\hline
26 &(0;2,0,13)&(0;2,0,5)&(0;0,0,0)&(1;0,4,2)&(0;0,1,0)&(0;1,0,0)&(0;0,0,0)&(0;0,2,0) & 17
\\
\hline
27 &(0;2,2,8)&(0;2,1,2)&(0;0,1,2)&(1;0,2,4)&(0;0,0,0)&(0;0,0,0)&(0;1,0,0)&(0;0,2,0) & 18
\\
\hline
28 &(0;2,3,8)&(0;1,1,3)&(0;1,0,1)&(1;0,3,3)&(0;0,1,0)&(0;0,0,0)&(0;1,0,1)&(0;0,1,0) & 19
\\
\hline
29 &(0;2,3,8)&(0;1,1,2)&(0;1,0,2)&(1;0,3,3)&(0;0,1,0)&(0;0,0,0)&(0;1,0,0)&(0;0,1,1) & 19
\\
\hline
30 &(0;2,6,3)&(0;0,2,1)&(0;2,0,1)&(1;0,2,4)&(0;0,1,1)&(0;0,0,0)&(0;0,0,1)&(0;1,0,1) & 21
\\
\hline
31 &(0;2,6,3)&(0;0,2,0)&(0;2,0,2)&(1;0,2,4)&(0;0,1,1)&(0;0,0,0)&(0;0,0,0)&(0;1,0,2) & 21
\\
\hline
32 &(0;5,0,0)&(0;0,0,0)&(0;0,0,5)&(0;0,5,0)&(1;0,0,0)&(0;0,0,0)&(0;0,0,0)&(0;0,0,5) & 15
\\
\hline
33 &(0;5,0,2)&(0;0,0,2)&(0;0,0,3)&(0;0,4,1)&(1;0,0,0)&(0;0,0,0)&(0;0,0,2)&(0;0,1,2) & 15
\\
\hline
34 &(0;5,0,4)&(0;0,0,4)&(0;0,0,1)&(0;0,3,2)&(1;0,0,0)&(0;0,0,1)&(0;0,0,2)&(0;0,2,0) & 15
\\
\hline
35 &(0;0,10,0)&(0;0,5,0)&(0;0,0,5)&(0;5,0,0)&(0;0,0,0)&(1;0,0,0)&(0;0,0,5)&(0;0,0,0) & 25
\\
\hline
36 &(0;0,10,0)&(0;0,5,0)&(0;0,0,4)&(0;5,0,0)&(0;0,0,1)&(1;0,0,0)&(0;0,0,3)&(0;0,0,2) & 25
\\
\hline
37 &(0;0,10,0)&(0;0,5,0)&(0;0,0,3)&(0;5,0,0)&(0;0,0,2)&(1;0,0,0)&(0;0,0,1)&(0;0,0,4) & 25
\\
\hline
38 &(0;2,6,2)&(0;1,2,2)&(0;0,2,1)&(0;3,2,2)&(0;0,0,2)&(0;0,0,1)&(1;0,0,0)&(0;0,0,2) & 24
\\
\hline
39 &(0;4,2,5)&(0;2,0,3)&(0;0,2,1)&(0;1,4,2)&(0;0,1,1)&(0;0,0,1)&(0;0,0,1)&(1;0,0,0) & 23
\\
\hline
40 &(0;4,2,5)&(0;2,0,2)&(0;0,2,2)&(0;1,4,2)&(0;0,1,1)&(0;0,0,0)&(0;0,0,2)&(1;0,0,0) & 23
\\
\hline\hline
\end{tabular*}
\label{T:vertexR3}
\end{table}
\begin{figure}
\centering
\includegraphics[width=0.67\linewidth]{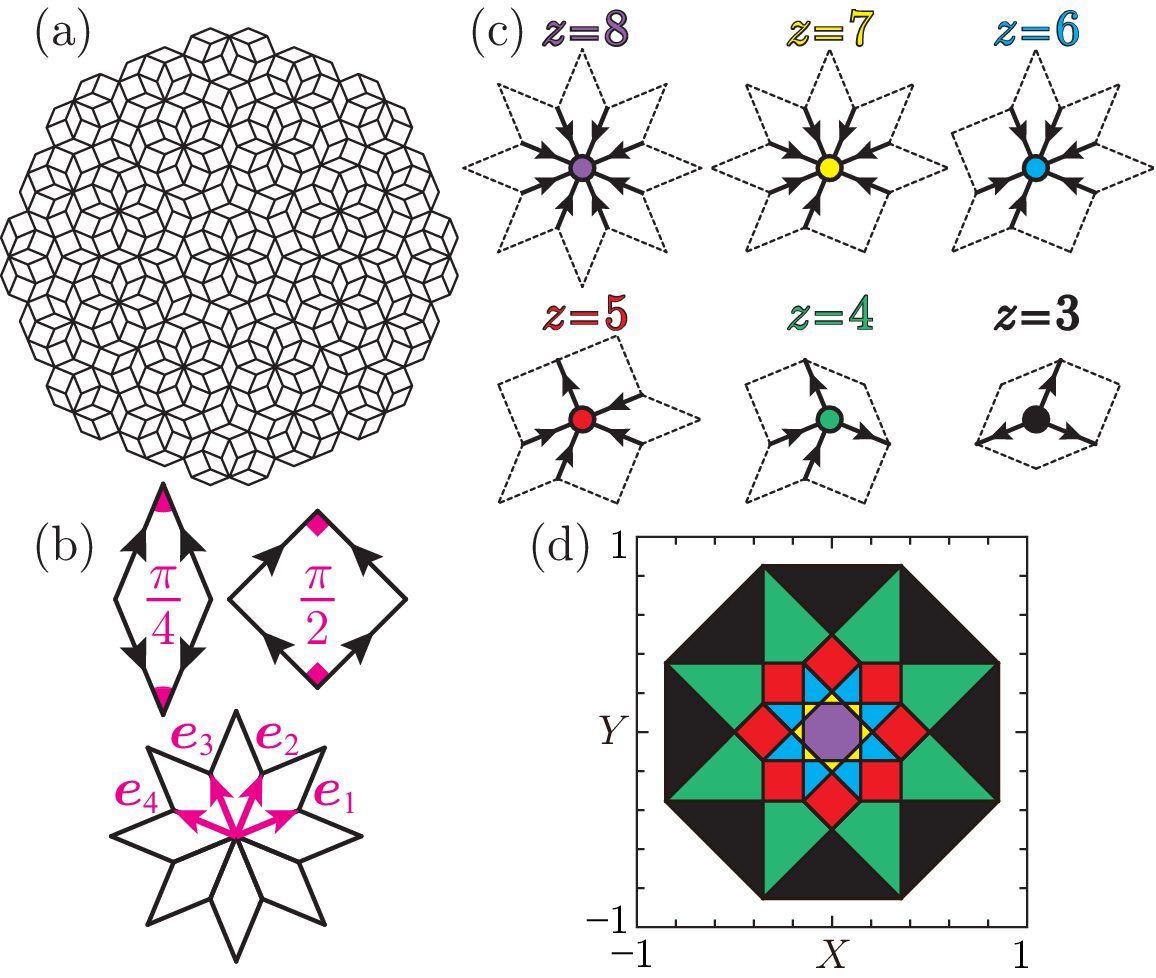}
\caption{%
         (a) A two-dimensional Ammann-Beenker lattice of $L=481$ with fourfold rotational symmetry.
         (b) The Ammann-Beenker tiling constitutes of a rhombus with angle $\frac{\pi}{4}$ and
             a square, whose edges each are marked with an arrow so as to define
             the matching rules.
             The canonical basis vectors of a four-dimensional hypercubic lattice convert into
             four vectors which we shall denote by
             $\bm{e}_{1}$, $\bm{e}_{2}$, $\bm{e}_{3}$, and $\bm{e}_{4}$.
             They are the primitive translation vectors for the two-dimensional
             Ammann-Beenker tiling.
         (c) Six types of local environments on the Ammann-Beenker lattice
             with their coordination numbers $z$ ranging from $3$ to $8$.
         (d) The perpendicular space of the Ammann-Beenker lattice is a single octagon,
             which is divided into several domains colored differently, each corresponding to
             any of the six types of local environments in the physical space.}
\label{F:AB}
\end{figure}

   The orthogonal complement of the two-dimensional Penrose lattice is a stack of four pentagons
[Figure \ref{F:Penrose}(d)].
The Penrose lattice is bipartite and its two sublattices reside separately in the pentagons at
$\sqrt{5}Z=0,2$ and those at $\sqrt{5}Z=1,3$ in the perpendicular space.
The infinite Penrose lattice consists of eight types of vertices with $z$ ranging from
$3\equiv z_{\mathrm{min}}$ to $7\equiv z_{\mathrm{max}}$ [Figure \ref{F:Penrose}(c)].
These eight types of vertices each fill certain compact regions inside the pentagons in
the perpendicular space, as is shown in Figure \ref{F:Penrose}(d).
The thus labeled domains are further subdivided, i.e., the eight species of vertices are further
subclassified, if we look over their surroundings to a longer distance \cite{G224201}.
Let us define a distance between any two vertices by counting the minimum number of bonds needed
to connect them and denote it by $R$.
Suppose we characterize any vertex as a function of $R$.
At the smallest $R$ that we shall define as $0$, every vertex is characterized by
its own coordination number only without seeing any surrounding vertex, resulting in
the naivest classification consisting of the eight species specified in Figure \ref{F:Penrose}(c).
These largest classes divide into more and more subclasses as we go farther and farther away from
the vertices of interest, i.e., with increasing $R$.
When we associate every vertex with its nearest neighbors, i.e., with its surrounding environment
truncated at $R=1$, the D and Q species, for instance, further divide into four and two subtypes,
respectively, as is illustrated with Figure \ref{F:local_exDandQ}.
Figure \ref{F:Penrose}(d) shows such subclassifications with increasing $R$ as well as the naivest
classification at $R=0$ in the perpendicular space.
The most basic $8$ species divide into $15$, $27$, and $40$ subtypes
when $R$ is set equal to $1$, $2$, and $3$, respectively.

   We take a glance at another two-dimensional quasiperiodic tiling for reference.
Figure \ref{F:AB}(a) is an Ammann-Beenker lattice in two dimensions, which can be generated
from minimal seeds [Figure \ref{F:AB}(b)] by matching up their arrowed edges so as to point
in the same direction and repeating the inflation-deflation operation \cite{K115125} with
a magnification ratio of the silver number $1+\sqrt{2}$ to unity.
Such operations yield the self-similarity of the Ammann-Beenker lattice.
Every two vertices on the Ammann-Beenker lattice are connected by a linear combination
of four independent primitive lattice vectors [Figure \ref{F:AB}(b)] with a unique set of
four integral coefficients.
The physical dimension $d$ and rank $D$ of the Ammann-Beenker lattice equal $2$ and $4$,
respectively \cite{L313}.
The Ammann-Beenker lattice is obtained as a projection of a four-dimensional hypercubic lattice
onto a two-dimensional physical space,
\begin{align}
   \left[
   \begin{array}{c}
    x \\
    y \\
    X \\
    Y
   \end{array}
   \right]
  =
   \frac{1}{\sqrt{2}}
   \left[
   \begin{array}{cccc}
    \cos\phi_{0} & \cos\phi_{1} & \cos\phi_{2} & \cos\phi_{3} \\
    \sin\phi_{0} & \sin\phi_{1} & \sin\phi_{2} & \sin\phi_{3} \\
    \cos\phi_{0} & \cos\phi_{3} & \cos\phi_{6} & \cos\phi_{9} \\
    \sin\phi_{0} & \sin\phi_{3} & \sin\phi_{6} & \sin\phi_{9} \\
   \end{array}
   \right]
   \left[
   \begin{array}{c}
    m_{1} \\
    m_{2} \\
    m_{3} \\
    m_{4}
   \end{array}
   \right]
\ \ \ 
   \left(
   \phi_{n}
   \equiv
    \frac{\pi}{4}n,\ 
   m_{n}\in\mathbb{Z}
   \right),
   \label{E:projectionAB}
\end{align}
where $(x,y)$ and $(X,Y)$ are the coordinates in the two-dimensional physical and
perpendicular spaces, respectively, and
the prefactor $1/\sqrt{2}$ plays a lattice constant in the former.
The points in the former and latter also have a one-to-one correspondence between them.
The orthogonal complement of the two-dimensional Ammann-Beenker lattice is a single octagon
[Figure \ref{F:AB}(d)].
The Ammann-Beenker lattice is also bipartite but its two sublattices are no longer separable
in the perpendicular space in contrast to those of the Penrose lattice.
The two sublattices of the Penrose lattice each assemble into a different couple of pentagons
in the perpendicular space, whereas those of the Ammann-Beenker lattice each form an identical
octagon in the perpendicular space in the thermodynamic limit.
The infinite Ammann-Beenker lattice consists of six types of vertices with $z$ ranging from
$3\equiv z_{\mathrm{min}}$ to $8\equiv z_{\mathrm{max}}$ [Figure \ref{F:AB}(c)].
These six types of vertices each fill certain compact regions inside the octagon in
the perpendicular space, as is shown in Figure \ref{F:AB}(d).

\section{Spin-Wave Formalism}
   We set the spin-$S$ nearest-neighbor antiferromagnetic Heisenberg model
\begin{align}
      \mathcal{H}
  =J\sum_{\langle i,j \rangle}
   \bm{S}_{\bm{r}_{i}}\cdot\bm{S}_{\bm{r}_{j}}
   \ 
   (J>0),
   \label{E:Ham}
\end{align}
on the two-dimensional $\mathbf{C}_\mathrm{5v}$ Penrose and $\mathbf{C}_\mathrm{8v}$ Ammann-Beenker
lattices, each with $L\equiv L_{\mathrm{A}}+L_{\mathrm{B}}$ sites, where
$\bm{S}_{\bm{r}_{i}}\,(i=1,2,\cdots,L_{\mathrm{A}})$ are the vector spin operators of magnitude
$S$ attached to the site at $\bm{r}_{i}$ forming a sublattice which we shall refer to as A, while
$\bm{S}_{\bm{r}_{j}}\,(j=1,2,\cdots,L_{\mathrm{B}})$ are the vector spin operators of magnitude
$S$ attached to the site at $\bm{r}_{j}$ forming another sublattice which we shall refer to as B.
We denote an arbitrary---in the sense of possibly running over both sublattices---site by
$\bm{r}_{l}$ in the following.
Suppose that $\langle i,j\rangle$ runs over all and only nearest neighbors.
We express the Hamiltonian \eqref{E:Ham} in terms of the Holstein-Primakoff bosonic spin deviation
operators \cite{H1098},
\begin{align}
   &
   S_{\bm{r}_{i}}^{+}
  =\left(2S-a_{i}^{\dagger}a_{i}\right)^{\frac{1}{2}}
   a_{i},\ 
   S_{\bm{r}_{i}}^{-}
  =a_{i}^{\dagger}
   \left(2S-a_{i}^{\dagger}a_{i}\right)^{\frac{1}{2}},\ 
   S_{\bm{r}_{i}}^{z}
  =S-a_{i}^{\dagger}a_{i};\ 
   \allowdisplaybreaks
   \nonumber \\
   &
   S_{\bm{r}_{j}}^{+}
  =b_{j}^{\dagger}
   \left(2S-b_{j}^{\dagger}b_{j}\right)^{\frac{1}{2}},\ 
   S_{\bm{r}_{j}}^{-}
  =\left(2S-b_{j}^{\dagger}b_{j}\right)^{\frac{1}{2}}
   b_{j},\ 
   S_{\bm{r}_{j}}^{z}
  =b_{j}^{\dagger}b_{j}-S,
   \label{E:HPboson}
\end{align}
and expand it in descending powers of the spin magnitude \cite{Y11033},
\begin{align}
    \mathcal{H}
   = \mathcal{H}^{(2)}
    +\mathcal{H}^{(1)}
    +\mathcal{H}^{(0)}
    +O(S^{-1}),
\end{align}
where $\mathcal{H}^{(m)}$, on the order of $S^{m}$, reads
\begin{align}
   &
   \mathcal{H}^{(2)}
 =-JS^2
   \sum_{i\in\mathrm{A}}\sum_{j\in\mathrm{B}}
   l_{i,j},\ 
   \mathcal{H}^{(1)}
  =JS
   \sum_{i\in\mathrm{A}}\sum_{j\in\mathrm{B}}
   l_{i,j}
   \left(
    a_{i}^{\dagger}a_{i}+b_{j}^{\dagger}b_{j}
   +a_{i}b_{j}+a_{i}^{\dagger}b_{j}^{\dagger}
   \right),
  \allowdisplaybreaks
  \nonumber \\
   &
   \mathcal{H}^{(0)}
 =-J
   \sum_{i\in\mathrm{A}}\sum_{j\in\mathrm{B}}
   l_{i,j}
   \left[
    a_{i}^{\dagger}a_{i}b_{j}^{\dagger}b_{j}
   +\frac{1}{4}
    \left(
     a_{i}^{\dagger}a_{i}a_{i}b_{j}
    +a_{i}^{\dagger}b_{j}^{\dagger}b_{j}^{\dagger}b_{j}
    +\mathrm{H.c.}
    \right)
   \right],
   \label{E:1Sexpansion}
\end{align}
with $l_{i,j}$ being $1$ for connected vertices $\bm{r}_{i}$ and $\bm{r}_{j}$, otherwise $0$.
We decompose the $O(S^{0})$ quartic Hamiltonian $\mathcal{H}^{(0)}$ into quadratic terms
$\mathcal{H}^{(0)}_{\mathrm{BL}}$ and quartic terms $:\mathcal{H}^{(0)}:$ through Wick's theorem
\cite{I053701,N034714,Y094412},
\begin{align}
   a_{i}^{\dagger}a_{i}b_{j}^{\dagger}b_{j}
  &=
   :a_{i}^{\dagger}a_{i}b_{j}^{\dagger}b_{j}:
  +{}^{\ [0]}\!\langle 0| b_{j}^{\dagger}b_{j} |0 \rangle^{\![0]}
   a_{i}^{\dagger}a_{i}
  +{}^{\ [0]}\!\langle 0| a_{i}^{\dagger}a_{i} |0 \rangle^{\![0]}
   b_{j}^{\dagger}b_{j}
  +{}^{\ [0]}\!\langle 0| a_{i}^{\dagger}b_{j}^{\dagger} |0 \rangle^{\![0]}
   a_{i}b_{j}
   \allowdisplaybreaks
   \nonumber \\
   &
  +{}^{\ [0]}\!\langle 0| a_{i}b_{j} |0 \rangle^{\![0]}
   a_{i}^{\dagger}b_{j}^{\dagger}
  -{}^{\ [0]}\!\langle 0| a_{i}^{\dagger}a_{i} |0 \rangle^{\![0]}
   {}^{\ [0]}\!\langle 0| b_{j}^{\dagger}b_{j} |0 \rangle^{\![0]}
  -{}^{\ [0]}\!\langle 0| a_{i}^{\dagger}b_{j}^{\dagger} |0 \rangle^{\![0]}
   {}^{\ [0]}\!\langle 0| a_{i}b_{j} |0 \rangle^{\![0]},
   \allowdisplaybreaks
   \nonumber \\
   a_{i}^{\dagger}a_{i}a_{i}b_{j}
  &=
   :a_{i}^{\dagger}a_{i}a_{i}b_{j}:
   \allowdisplaybreaks
   \nonumber \\
   &
  +2\left(
     {}^{\ [0]}\!\langle 0| a_{i}^{\dagger}a_{i} |0 \rangle^{\![0]}
     a_{i}b_{j}
    +{}^{\ [0]}\!\langle 0| a_{i}b_{j} |0 \rangle^{\![0]}
     a_{i}^{\dagger}a_{i}
    -{}^{\ [0]}\!\langle 0| a_{i}^{\dagger}a_{i} |0 \rangle^{\![0]}
     {}^{\ [0]}\!\langle 0| a_{i}b_{j} |0 \rangle^{\![0]}
   \right),
   \allowdisplaybreaks
   \nonumber \\
   a_{i}^{\dagger}b_{j}^{\dagger}b_{j}^{\dagger}b_{j}
  &=
   :a_{i}^{\dagger}b_{j}^{\dagger}b_{j}^{\dagger}b_{j}:
   \allowdisplaybreaks
   \nonumber \\
   &
  +2\left(
     {}^{\ [0]}\!\langle 0| a_{i}^{\dagger}b_{j}^{\dagger} |0 \rangle^{\![0]}
     b_{j}^{\dagger}b_{j}
    +{}^{\ [0]}\!\langle 0| b_{j}^{\dagger}b_{j} |0 \rangle^{\![0]}
     a_{i}^{\dagger}b_{j}^{\dagger}
    -{}^{\ [0]}\!\langle 0| a_{i}^{\dagger}b_{j}^{\dagger} |0 \rangle^{\![0]}
     {}^{\ [0]}\!\langle 0| b_{j}^{\dagger}b_{j} |0 \rangle^{\![0]}
   \right),
   \label{E:Wickdecomp}
\end{align}
where every normal ordering $:\cdots:$ is based on the up-to-$O(S^0)$ magnon operators
$\alpha_{k_\sigma}^{\sigma[0]\dagger}$ and $\alpha_{k_\sigma}^{\sigma[0]}$
[cf. \eqref{E:HBLdiag}],
just like
$
   :\alpha_{k_{+}}^{+[0]}\alpha_{k_{+}'}^{+[0]\dagger}
    \alpha_{k_{-}}^{-[0]}\alpha_{k_{-}'}^{-[0]\dagger}:
  =
   \alpha_{k_{+}'}^{+[0]\dagger}\alpha_{k_{-}'}^{-[0]\dagger}
   \alpha_{k_{+}}^{+[0]}\alpha_{k_{-}}^{-[0]}
$,
and their vacuum state is denoted by $|0 \rangle^{\![0]}$.
We thus define the up-to-$O(S^0)$ bilinear interacting SW (ISW) Hamiltonian \cite{Y064426}
\begin{align}
   \mathcal{H}_{\mathrm{BL}}^{[0]}
  \equiv
   \mathcal{H}^{(2)}
  +\mathcal{H}^{(1)}
  +\mathcal{H}^{(0)}_{\mathrm{BL}}
   \label{E:HBL[0]}
\end{align}
as well as the up-to-$O(S^1)$ linear SW (LSW) Hamiltonian
\begin{align}
   \mathcal{H}_{\mathrm{BL}}^{[1]}
  \equiv
   \mathcal{H}^{(2)}
  +\mathcal{H}^{(1)}
   \label{E:HBL[1]}
\end{align}
whose ground state is denoted by $|0 \rangle^{\![1]}$.

   Let us introduce row vectors $\bm{a}^{\dagger}$ and $\bm{b}^{\dagger}$, of dimension
$L_{\mathrm{A}}$ and $L_{\mathrm{B}}$, respectively, as
\begin{align}
   \bm{c}^{\dagger}
  =\left[
    a_{1}^{\dagger}, \cdots, a_{L_{\mathrm{A}}}^{\dagger},
    b_{1},\cdots, b_{L_{\mathrm{B}}}
   \right]
  \equiv
   \left[\bm{a}^{\dagger}, {}^{t}\bm{b}\right]
   \label{E:cVector}
\end{align}
and matrices $\mathbf{A}^{[m]}$, $\mathbf{B}^{[m]}$, and $\mathbf{C}^{[m]}$
relevant to the up-to-$O(S^m)$ approximation, of dimension
$L_{\mathrm{A}}\times L_{\mathrm{A}}$,
$L_{\mathrm{B}}\times L_{\mathrm{B}}$, and
$L_{\mathrm{A}}\times L_{\mathrm{B}}$, respectively, as
\begin{align}
   &
   \left[\mathbf{A}^{[1]}\right]_{i,i'}
  =\delta_{i,i'}J\sum_{j\in\mathrm{B}}l_{i,j}S
  =\delta_{i,i'}z_iSJ,\ 
   \left[\mathbf{B}^{[1]}\right]_{j,j'}
  =\delta_{j,j'}J\sum_{i\in\mathrm{A}}l_{i,j}S
  =\delta_{j,j'}z_jSJ,\ 
   \left[\mathbf{C}^{[1]}\right]_{i,j}
  =l_{i,j}SJ,
   \label{E:ABCmatrices[1]}
   \allowdisplaybreaks
   \\
   &
   \left[\mathbf{A}^{[0]}\right]_{i,i'}
  =\delta_{i,i'}J\sum_{j\in\mathrm{B}}l_{i,j}
   \left[
     S
    -{}^{\ [0]}\!\langle 0| b_{j}^{\dagger}b_{j} |0 \rangle^{\![0]}
    -\frac{1}{2}\left(
        {}^{\ [0]}\!\langle 0| a_{i}b_{j} |0 \rangle^{\![0]}
       +{}^{\ [0]}\!\langle 0| a_{i}^{\dagger}b_{j}^{\dagger} |0 \rangle^{\![0]}
     \right)
   \right],
   \allowdisplaybreaks
   \nonumber \\
   &
   \left[\mathbf{B}^{[0]}\right]_{j,j'}
  =\delta_{j,j'}J\sum_{i\in\mathrm{A}}l_{i,j}
   \left[
     S
    -{}^{\ [0]}\!\langle 0| a_{i}^{\dagger}a_{i} |0 \rangle^{\![0]}
    -\frac{1}{2}\left(
        {}^{\ [0]}\!\langle 0| a_{i}b_{j} |0 \rangle^{\![0]}
       +{}^{\ [0]}\!\langle 0| a_{i}^{\dagger}b_{j}^{\dagger} |0 \rangle^{\![0]}
     \right)
   \right],
   \allowdisplaybreaks
   \nonumber \\
   &
   \left[\mathbf{C}^{[0]}\right]_{i,j}
  =l_{i,j}J\left[
     S
    -{}^{\ [0]}\!\langle 0| a_{i}b_{j} |0 \rangle^{\![0]}
    -\frac{1}{2}\left(
        {}^{\ [0]}\!\langle 0| a_{i}^{\dagger}a_{i} |0 \rangle^{\![0]}
       +{}^{\ [0]}\!\langle 0| b_{j}^{\dagger}b_{j} |0 \rangle^{\![0]}
     \right)
   \right],
   \label{E:ABCmatrices[0]}
\end{align}
where $z_l$ is the coordination number of the vertex at $\bm{r}_l$.
Then we can compactly express the bilinear Hamiltonian, whether up to $O(S^1)$ or $O(S^0)$,
in a matrix notation as
\begin{align}
   \mathcal{H}_{\mathrm{BL}}^{[1]}
  =\mathcal{H}^{(2)}
  +\varepsilon^{(1)}
  +\bm{c}^{\dagger}\mathbf{M}^{[1]}\bm{c},\ 
   \mathcal{H}_{\mathrm{BL}}^{[0]}
  =\mathcal{H}^{(2)}
  +\varepsilon^{(1)}
  +\varepsilon^{(0)}
  +\bm{c}^{\dagger}\mathbf{M}^{[0]}\bm{c};\ \ 
   \mathbf{M}^{[m]}
   \equiv
   \left[
   \begin{array}{c|c}
    \!\mathbf{A}^{[m]}\!\!\!           & \mathbf{C}^{[m]}\!\!\!
   \\ \hline
    \!\mathbf{C}^{[m]\dagger}\!\!\! & \mathbf{B}^{[m]}\!\!\!
   \end{array}
   \right],
   \label{E:BLH1}
\end{align}
where the $O(S^m)$ constants $\varepsilon^{(m)}$ are given by
\begin{align}
   &
   \varepsilon^{(1)}
 =-JS
   \sum_{i\in\mathrm{A}}\sum_{j\in\mathrm{B}}
   l_{i,j},
   \allowdisplaybreaks
   \nonumber \\
   &
   \varepsilon^{(0)}
  =J\sum_{i\in\mathrm{A}}\sum_{j\in\mathrm{B}}
   l_{i,j}
   \biggl[
     {}^{\ [0]}\!\langle 0| a_{i}^{\dagger}a_{i} |0 \rangle^{\![0]}
    +\frac{1}{2}
     \left(
       {}^{\ [0]}\!\langle 0| a_{i}b_{j} |0 \rangle^{\![0]}
      +{}^{\ [0]}\!\langle 0| a_{i}^{\dagger}b_{j}^{\dagger} |0 \rangle^{\![0]}
     \right)
   \allowdisplaybreaks
   \nonumber \\
   &\quad\ \ 
    +{}^{\ [0]}\!\langle 0| a_{i}^{\dagger}a_{i} |0 \rangle^{\![0]}
     {}^{\ [0]}\!\langle 0| b_{j}^{\dagger}b_{j} |0 \rangle^{\![0]}
    +{}^{\ [0]}\!\langle 0| a_{i}^{\dagger}b_{j}^{\dagger} |0 \rangle^{\![0]}
     {}^{\ [0]}\!\langle 0| a_{i}b_{j} |0 \rangle^{\![0]}
   \allowdisplaybreaks
   \nonumber \\
   &\quad\ \ 
    +\frac{1}{2}
     \left(
       {}^{\ [0]}\!\langle 0| a_{i}^{\dagger}a_{i} |0 \rangle^{\![0]}
      +{}^{\ [0]}\!\langle 0| b_{j}^{\dagger}b_{j} |0 \rangle^{\![0]}
     \right)
     \left(
       {}^{\ [0]}\!\langle 0| a_{i}^{\dagger}b_{j}^{\dagger} |0 \rangle^{\![0]}
      +{}^{\ [0]}\!\langle 0| a_{i}b_{j} |0 \rangle^{\![0]}
     \right)
    \biggr].
   \label{E:EBL}
\end{align}
We carry out the Bogoliubov transformation
\begin{align}
   &
   \bm{c}
  =\left[
   \begin{array}{c|c}
    \mathbf{S}^{[m]} & \mathbf{U}^{[m]} \\ \hline
    \mathbf{V}^{[m]} & \mathbf{T}^{[m]}
   \end{array}
   \right]
   \bm{\alpha}^{[m]};\ \ 
   \left[\mathbf{S}^{[m]}\right]_{i,k_{-}}\!\!
  \equiv
   s_{i,k_{-}}^{[m]},\ 
   \left[\mathbf{T}^{[m]}\right]_{j,k_{+}}\!\!
  \equiv
   t_{j,k_{+}}^{[m]},\ 
   \left[\mathbf{U}^{[m]}\right]_{i,k_{+}}\!\!
  \equiv
   u_{i,k_{+}}^{[m]},\ 
   \left[\mathbf{V}^{[m]}\right]_{j,k_{-}}\!\!
  \equiv
   v_{j,k_{-}}^{[m]}
   \label{E:Bogoliubov}
\end{align}
with the matrices
$\mathbf{S}^{[m]}$, $\mathbf{T}^{[m]}$, $\mathbf{U}^{[m]}$, and $\mathbf{V}^{[m]}$,
of dimension
$L_{\mathrm{A}}\times L_{-}$,
$L_{\mathrm{B}}\times L_{+}$,
$L_{\mathrm{A}}\times L_{+}$, and
$L_{\mathrm{B}}\times L_{-}$, respectively,
to obtain the quasiparticle magnons
\begin{align}
   \left[\alpha_{1}^{-[m]\dagger},\cdots,\alpha_{L_{-}}^{-[m]\dagger},
         \alpha_{1}^{+[m]},\cdots,\alpha_{L_{+}}^{+[m]}\right]
  \equiv
   \bm{\alpha}^{[m]\dagger}
   \label{E:magnondef}
\end{align}
and their Hamiltonian
\begin{align}
   \mathcal{H}_{\mathrm{BL}}^{[m]}
  =E^{[m]}
  +\sum_{\sigma=\mp}
   \sum_{k_{\sigma}=1}^{L_{\sigma}}
   \varepsilon_{k_{\sigma}}^{\sigma[m]}
   \alpha_{k_{\sigma}}^{\sigma[m]\dagger}\alpha_{k_{\sigma}}^{\sigma[m]},
   \label{E:HBLdiag}
\end{align}
where $E^{[m]}$ is the up-to-$O(S^m)$ ground-state energy,
\begin{align}
   E^{[1]}
  =\mathcal{H}^{(2)}
  +\varepsilon^{(1)}
  +\sum_{k_{+}=1}^{L_{+}}
   \varepsilon_{k_{+}}^{+[1]},\ 
   E^{[0]}
  =\mathcal{H}^{(2)}
  +\varepsilon^{(1)}
  +\varepsilon^{(0)}
  +\sum_{k_{+}=1}^{L_{+}}
   \varepsilon_{k_{+}}^{+[0]},
\end{align}
while $\alpha_{k_\sigma}^{\sigma[m]\dagger}$ creates an up-to-$O(S^m)$ ferromagnetic ($\sigma=-$)
or antiferromagnetic ($\sigma=+$) magnon of energy $\varepsilon_{k_{\sigma}}^{\sigma[m]}$,
reducing or enhancing the ground-state magnetization \cite{B3921,YR14008}, respectively.
Unless $L\rightarrow\infty$, $L_-$ equals $L_{\mathrm{A}}$ minus the number of zero
eigenvalues, i.e., $L_{\mathrm{A}}-1$, while $L_+$ equals $L_{\mathrm{B}}$.
In the thermodynamic limit $L\rightarrow\infty$ with $L_{\mathrm{A}}=L_{\mathrm{B}}=L/2$,
$L_{\mp}$ both equal $L/2-1$.
Now that we have obtained \eqref{E:HBLdiag}, a Baker-Campbell-Hausdorff relation \cite{L101}
immediately shows that the Heisenberg magnon operators evolve in time as
\begin{align}
   \alpha_{k_{\sigma}}^{\sigma[m]}(t)
  \equiv
   e^{\frac{i}{\hbar}\mathcal{H}_{\mathrm{BL}}^{[m]}t}
   \alpha_{k_{\sigma}}^{\sigma[m]}
   e^{-\frac{i}{\hbar}\mathcal{H}_{\mathrm{BL}}^{[m]}t}
  =e^{-\frac{i}{\hbar}\varepsilon_{k_{\sigma}}^{\sigma[m]}t}
   \alpha_{k_{\sigma}}^{\sigma[m]}.
   \label{E:BCHrelation}
\end{align}

\section{Spin-Wave Dynamics}
   We define the up-to-$O(S^m)$ dynamic structure factors
\begin{align}
   &
   S^{\lambda\lambda}(\bm{q};\omega)^{\![m]}
  \equiv
   \sum_{l,l'=1}^{L}
   \frac{e^{i\bm{q}\cdot(\bm{r}_{l}-\bm{r}_{l'})}}{2\pi\hbar L}
   \int_{-\infty}^{\infty}
   {}^{\ [m]}\!\langle 0|
    \delta S_{\bm{r}_{l}}^{\lambda}(t) \delta S_{\bm{r}_{l'}}^{\lambda}
   | 0 \rangle^{\![m]}
   e^{i\omega t}dt\ (\lambda=x,y,z)
   \label{E:SqwDefxyz}
\end{align}
in terms of the spin deviations from the up-to-$O(S^m)$ ground state
$
   \delta S_{\bm{r}_{l}}^{\lambda}
  \equiv
   S_{\bm{r}_{l}}^{\lambda}
  -{}^{\ [m]}\!\langle 0|
     S_{\bm{r}_{l}}^{\lambda}
   | 0 \rangle^{\![m]}\ 
$
rather than the projection values themselves.
Note that
$\delta S_{\bm{r}_{l}}^x=S_{\bm{r}_{l}}^x$ and
$\delta S_{\bm{r}_{l}}^y=S_{\bm{r}_{l}}^y$,
whether we employ LSWs or ISWs, in the present representation.
We refer to
$S^{xx}(\bm{q};\omega)^{\![m]}+S^{yy}(\bm{q};\omega)^{\![m]}
\equiv
 S^\perp(\bm{q};\omega)^{\![m]}$ and
$S^{zz}(\bm{q};\omega)^{\![m]}
\equiv
 S^\parallel(\bm{q};\omega)^{\![m]}$
as the transverse and longitudinal components, respectively.
The transverse contribution to the dynamic structure factor is more tractable in terms of the spin
raising and lowering operators.
Suppose we define
\begin{align}
   S^{\pm\mp}(\bm{q};\omega)^{\![m]}
  \equiv
   \sum_{l,l'=1}^{L}
   \frac{e^{i\bm{q}\cdot(\bm{r}_{l}-\bm{r}_{l'})}}{2\pi\hbar L}
   \int_{-\infty}^{\infty}
   {}^{\ [m]}\!\langle 0|
    S_{\bm{r}_{l}}^{\pm}(t)S_{\bm{r}_{l'}}^{\mp}
   | 0 \rangle^{\![m]}
   e^{i\omega t}dt;\ 
   S_{\bm{r}_{l}}^{\pm}
  =S_{\bm{r}_{l}}^{x}\pm iS_{\bm{r}_{l}}^{y},
   \label{E:SqwDeftransverse}
\end{align}
then we have
$
   S^\perp(\bm{q};\omega)^{\![m]}
  =[
    S^{+-}(\bm{q};\omega)^{\![m]}
   +S^{-+}(\bm{q};\omega)^{\![m]}
   ]/2
$.
In terms of the up-to-$O(S^m)$ magnon operators, the Fourier-transformed spin operators
\begin{align}
   S_{\bm{q}}^\pm
  \equiv
   \frac{1}{\sqrt{L}}\sum_{l=1}^{L}
   e^{i\bm{q}\cdot\bm{r}_{l}}
   S_{\bm{r}_{l}}^\pm,\ 
   \delta S_{\bm{q}}^z
  \equiv
   \frac{1}{\sqrt{L}}\sum_{l=1}^{L}
   e^{i\bm{q}\cdot\bm{r}_{l}}
   \delta S_{\bm{r}_{l}}^z
   \label{E:FT}
\end{align}
are expanded in descending powers of $S$ as
\begin{align}
   &
   S_{\bm{q}}^{+}
  =\left(
    S_{-\bm{q}}^{-}
   \right)^\dagger
  =\sum_{k_{-}=1}^{L_{-}}
   D_{-}^{(\frac{1}{2})}(\bm{q};k_{-})
   \alpha_{k_{-}}^{-[m]}
  +\sum_{k_{+}=1}^{L_{+}}
   C_{+}^{(\frac{1}{2})}(\bm{q};k_{+})
   \alpha_{k_{+}}^{+[m]\dagger}
   \allowdisplaybreaks
   \nonumber \\
   &\quad\ \qquad\qquad
  +\sum_{k_{-}=1}^{L_{-}}
   D_{-}^{(-\frac{1}{2})}(\bm{q};k_{-})
   \alpha_{k_{-}}^{-[m]}
  +\sum_{k_{+}=1}^{L_{+}}
   C_{+}^{(-\frac{1}{2})}(\bm{q};k_{+})
   \alpha_{k_{+}}^{+[m]\dagger}
   \allowdisplaybreaks
   \nonumber \\
   &\quad\ \qquad\qquad
  +\sum_{k_{-}=1}^{L_{-}}\sum_{k_{-}'=1}^{L_{-}}\sum_{k_{-}''=1}^{L_{-}}
   D_{---}^{(-\frac{1}{2})}(\bm{q};k_{-},k_{-}',k_{-}'')
   \alpha_{k_{-}}^{-[m]\dagger}\alpha_{k_{-}'}^{-[m]}\alpha_{k_{-}''}^{-[m]}
   \allowdisplaybreaks
   \nonumber \\
   &\quad\ \qquad\qquad
  +\sum_{k_{-}=1}^{L_{-}}\sum_{k_{+}'=1}^{L_{+}}\sum_{k_{+}''=1}^{L_{+}}
   D_{-++}^{(-\frac{1}{2})}(\bm{q};k_{-},k_{+}',k_{+}'')
   \alpha_{k_{-}}^{-[m]}\alpha_{k_{+}'}^{+[m]\dagger}\alpha_{k_{+}''}^{+[m]}
   \allowdisplaybreaks
   \nonumber \\
   &\quad\ \qquad\qquad
  +\sum_{k_{-}=1}^{L_{-}}\sum_{k_{-}'=1}^{L_{-}}\sum_{k_{+}''=1}^{L_{+}}
   D_{--+}^{(-\frac{1}{2})}(\bm{q};k_{-},k_{-}',k_{+}'')
   \alpha_{k_{-}}^{-[m]}\alpha_{k_{-}'}^{-[m]}\alpha_{k_{+}''}^{+[m]}
   \allowdisplaybreaks
   \nonumber \\
   &\quad\ \qquad\qquad
  +\sum_{k_{+}=1}^{L_{+}}\sum_{k_{+}'=1}^{L_{+}}\sum_{k_{+}''=1}^{L_{+}}
   C_{+++}^{(-\frac{1}{2})}(\bm{q};k_{+},k_{+}',k_{+}'')
   \alpha_{k_{+}}^{+[m]\dagger}\alpha_{k_{+}'}^{+[m]\dagger}\alpha_{k_{+}''}^{+[m]}
   \allowdisplaybreaks
   \nonumber \\
   &\quad\ \qquad\qquad
  +\sum_{k_{-}=1}^{L_{-}}\sum_{k_{-}'=1}^{L_{-}}\sum_{k_{+}''=1}^{L_{+}}
   C_{--+}^{(-\frac{1}{2})}(\bm{q};k_{-},k_{-}',k_{+}'')
   \alpha_{k_{-}}^{-[m]\dagger}\alpha_{k_{-}'}^{-[m]}\alpha_{k_{+}''}^{+[m]\dagger}
   \allowdisplaybreaks
   \nonumber \\
   &\quad\ \qquad\qquad
  +\sum_{k_{-}=1}^{L_{-}}\sum_{k_{+}'=1}^{L_{+}}\sum_{k_{+}''=1}^{L_{+}}
   C_{-++}^{(-\frac{1}{2})}(\bm{q};k_{-},k_{+}',k_{+}'')
   \alpha_{k_{-}}^{-[m]\dagger}\alpha_{k_{+}'}^{+[m]\dagger}\alpha_{k_{+}''}^{+[m]\dagger}
  +O(S^{-\frac{3}{2}}),
   \label{E:deltaSqT}
   \allowdisplaybreaks
   \\
   &
   \delta S_{\bm{q}}^z
  =\sum_{k_{-}=1}^{L_{-}}\sum_{k_{-}'=1}^{L_{-}}
   N_{--}^{(0)}(\bm{q};k_{-},k_{-}')
   \alpha_{k_{-}}^{-[m]\dagger}\alpha_{k_{-}'}^{-[m]}
  +\sum_{k_{+}=1}^{L_{+}}\sum_{k_{+}'=1}^{L_{+}}
   N_{++}^{(0)}(\bm{q};k_{+},k_{+}')
   \alpha_{k_{+}}^{+[m]\dagger}\alpha_{k_{+}'}^{+[m]}
   \allowdisplaybreaks
   \nonumber \\
   &\quad\ \ \ 
  +\sum_{k_{-}=1}^{L_{-}}\sum_{k_{+}'=1}^{L_{+}}
   N_{-+}^{(0)}(\bm{q};k_{-},k_{+}')
   \alpha_{k_{-}}^{-[m]}\alpha_{k_{+}'}^{+[m]}
  +\sum_{k_{-}=1}^{L_{-}}\sum_{k_{+}'=1}^{L_{+}}
   N_{-+}^{(0)}(-\bm{q};k_{-},k_{+}')^{*}
   \alpha_{k_{-}}^{-[m]\dagger}\alpha_{k_{+}'}^{+[m]\dagger},
   \label{E:deltaSqL}
\end{align}
where the coefficients
$D_{\sigma}^{(m)}(\bm{q};k_{\sigma})$ and
$C_{\sigma}^{(m)}(\bm{q};k_{\sigma})$ of order $S^m$,
$D_{\sigma\sigma'\sigma''}^{(-\frac{1}{2})}(\bm{q};k_{\sigma},k_{\sigma'}',k_{\sigma''}'')$ and
$C_{\sigma\sigma'\sigma''}^{(-\frac{1}{2})}(\bm{q};k_{\sigma},k_{\sigma'}',k_{\sigma''}'')$
of order $S^{(-\frac{1}{2})}$, and
$N_{\sigma\sigma'}^{(0)}(\bm{q};k_{\sigma},k_{\sigma'}')$ of order $S^0$
can be written in terms of the Bogoliubov transformation matrix \eqref{E:Bogoliubov},
\begin{align}
   &
   D_{-}^{(\frac{1}{2})}(\bm{q};k_{-})
  \equiv
   \sum_{l=1}^{L}
   \tilde{D}_{-}^{(\frac{1}{2})}(l,\bm{q};k_{-})
  =\frac{\sqrt{2S}}{\sqrt{L}}
   \left(
    \sum_{i\in\mathrm{A}}e^{i\bm{q}\cdot\bm{r}_{i}}
    s_{i,k_{-}}^{[m]}
   +\sum_{j\in\mathrm{B}}e^{i\bm{q}\cdot\bm{r}_{j}}
    v_{j,k_{-}}^{[m]}
   \right),
   \nonumber
   \allowdisplaybreaks
   \\
   &
   C_{+}^{(\frac{1}{2})}(\bm{q};k_{+})
  \equiv
   \sum_{l=1}^{L}
   \tilde{C}_{+}^{(\frac{1}{2})}(l,\bm{q};k_{+})
  =\frac{\sqrt{2S}}{\sqrt{L}}
   \left(
    \sum_{i\in\mathrm{A}}e^{i\bm{q}\cdot\bm{r}_{i}}
    u_{i,k_{+}}^{[m]}
   +\sum_{j\in\mathrm{B}}e^{i\bm{q}\cdot\bm{r}_{j}}
    t_{j,k_{+}}^{[m]}
   \right),
   \nonumber
   \allowdisplaybreaks
   \\
   &
   D_{-}^{(-\frac{1}{2})}(\bm{q};k_{-})
  \equiv
   \sum_{l=1}^{L}
   \tilde{D}_{-}^{(-\frac{1}{2})}(l,\bm{q};k_{-})
   \nonumber
   \allowdisplaybreaks
   \\
   &\qquad\qquad\quad\ 
  =-\frac{\sqrt{2S}}{2S\sqrt{L}}
   \left(
    \sum_{i\in\mathrm{A}}e^{i\bm{q}\cdot\bm{r}_{i}}
    \sum_{k_{+}'=1}^{L_+}\left|u_{i,k_{+}'}^{[m]}\right|^{2}
    s_{i,k_{-}}^{[m]}
   +\sum_{j\in\mathrm{B}}e^{i\bm{q}\cdot\bm{r}_{j}}
    \sum_{k_{-}'=1}^{L_-}\left|v_{j,k_{-}'}^{[m]}\right|^{2}
    v_{j,k_{-}}^{[m]}
   \right),
   \nonumber
   \allowdisplaybreaks
   \\
   &
   C_{+}^{(-\frac{1}{2})}(\bm{q};k_{+})
  \equiv
   \sum_{l=1}^{L}
   \tilde{C}_{+}^{(-\frac{1}{2})}(l,\bm{q};k_{+})
   \nonumber
   \allowdisplaybreaks
   \\
   &\qquad\qquad\quad\ 
  =-\frac{\sqrt{2S}}{2S\sqrt{L}}
   \left(
    \sum_{i\in\mathrm{A}}e^{i\bm{q}\cdot\bm{r}_{i}}
    \sum_{k_{+}'=1}^{L_+}\left|u_{i,k_{+}'}^{[m]}\right|^{2}
    u_{i,k_{+}}^{[m]}
   +\sum_{j\in\mathrm{B}}e^{i\bm{q}\cdot\bm{r}_{j}}
    \sum_{k_{-}'=1}^{L_-}\left|v_{j,k_{-}'}^{[m]}\right|^{2}
    t_{j,k_{+}}^{[m]}
   \right),
   \nonumber
   \allowdisplaybreaks
   \\
   &
   D_{---}^{(-\frac{1}{2})}(\bm{q};k_{-},k_{-}',k_{-}'')
  =-\frac{\sqrt{2S}}{4S\sqrt{L}}
   \left(
    \sum_{i\in\mathrm{A}}e^{i\bm{q}\cdot\bm{r}_{i}}
    s_{i,k_{-}}^{[m]*}s_{i,k_{-}'}^{[m]}s_{i,k_{-}''}^{[m]}
   +\sum_{j\in\mathrm{B}}e^{i\bm{q}\cdot\bm{r}_{j}}
    v_{j,k_{-}}^{[m]*}v_{j,k_{-}'}^{[m]}v_{j,k_{-}''}^{[m]}
   \right),
   \nonumber
   \allowdisplaybreaks
   \\
   &
   D_{-++}^{(-\frac{1}{2})}(\bm{q};k_{-},k_{+}',k_{+}'')
  =-\frac{\sqrt{2S}}{2S\sqrt{L}}
   \left(
    \sum_{i\in\mathrm{A}}e^{i\bm{q}\cdot\bm{r}_{i}}
    s_{i,k_{-}}^{[m]}u_{i,k_{+}'}^{[m]}u_{i,k_{+}''}^{[m]*}
   +\sum_{j\in\mathrm{B}}e^{i\bm{q}\cdot\bm{r}_{j}}
    v_{j,k_{-}}^{[m]}t_{j,k_{+}'}^{[m]}t_{j,k_{+}''}^{[m]*}
   \right),
   \nonumber
   \allowdisplaybreaks
   \\
   &
   D_{--+}^{(-\frac{1}{2})}(\bm{q};k_{-},k_{-}',k_{+}'')
  =-\frac{\sqrt{2S}}{4S\sqrt{L}}
   \left(
    \sum_{i\in\mathrm{A}}e^{i\bm{q}\cdot\bm{r}_{i}}
    s_{i,k_{-}}^{[m]}s_{i,k_{-}'}^{[m]}u_{i,k_{+}''}^{[m]*}
   +\sum_{j\in\mathrm{B}}e^{i\bm{q}\cdot\bm{r}_{j}}
    v_{j,k_{-}}^{[m]}v_{j,k_{-}'}^{[m]}t_{j,k_{+}''}^{[m]*}
   \right),
   \nonumber
   \allowdisplaybreaks
   \\
   &
   C_{+++}^{(-\frac{1}{2})}(\bm{q};k_{+},k_{+}',k_{+}'')
  =-\frac{\sqrt{2S}}{4S\sqrt{L}}
   \left(
    \sum_{i\in\mathrm{A}}e^{i\bm{q}\cdot\bm{r}_{i}}
    u_{i,k_{+}}^{[m]}u_{i,k_{+}'}^{[m]}u_{i,k_{+}''}^{[m]*}
   +\sum_{j\in\mathrm{B}}e^{i\bm{q}\cdot\bm{r}_{j}}
    t_{j,k_{+}}^{[m]}t_{j,k_{+}'}^{[m]}t_{j,k_{+}''}^{[m]*}
   \right),
   \nonumber
   \allowdisplaybreaks
   \\
   &
   C_{--+}^{(-\frac{1}{2})}(\bm{q};k_{-},k_{-}',k_{+}'')
  =-\frac{\sqrt{2S}}{2S\sqrt{L}}
   \left(
    \sum_{i\in\mathrm{A}}e^{i\bm{q}\cdot\bm{r}_{i}}
    s_{i,k_{-}}^{[m]*}s_{i,k_{-}'}^{[m]}u_{i,k_{+}''}^{[m]}
   +\sum_{j\in\mathrm{B}}e^{i\bm{q}\cdot\bm{r}_{j}}
    v_{j,k_{-}}^{[m]*}v_{j,k_{-}'}^{[m]}t_{j,k_{+}''}^{[m]}
   \right),
   \nonumber
   \allowdisplaybreaks
   \\
   &
   C_{-++}^{(-\frac{1}{2})}(\bm{q};k_{-},k_{+}',k_{+}'')
  =-\frac{\sqrt{2S}}{4S\sqrt{L}}
   \left(
    \sum_{i\in\mathrm{A}}e^{i\bm{q}\cdot\bm{r}_{i}}
    s_{i,k_{-}}^{[m]*}u_{i,k_{+}'}^{[m]}u_{i,k_{+}''}^{[m]}
   +\sum_{j\in\mathrm{B}}e^{i\bm{q}\cdot\bm{r}_{j}}
    v_{j,k_{-}}^{[m]*}t_{j,k_{+}'}^{[m]}t_{j,k_{+}''}^{[m]}
   \right),
   \nonumber
   \allowdisplaybreaks
   \\
   &
   N_{--}^{(0)}(\bm{q};k_{-},k_{-}')
  =\frac{1}{\sqrt{L}}
   \left(
    -\sum_{i\in\mathrm{A}}
     e^{i\bm{q}\cdot\bm{r}_{i}}
     s_{i,k_{-}}^{[m]*}s_{i,k_{-}'}^{[m]}
    +\sum_{j\in\mathrm{B}}
     e^{i\bm{q}\cdot\bm{r}_{j}}
     v_{j,k_{-}}^{[m]*}v_{j,k_{-}'}^{[m]}
   \right),
   \nonumber
   \allowdisplaybreaks
   \\
   &
   N_{++}^{(0)}(\bm{q};k_{+},k_{+}')
  =\frac{1}{\sqrt{L}}
   \left(
    -\sum_{i\in\mathrm{A}}
     e^{i\bm{q}\cdot\bm{r}_{i}}
     u_{i,k_{+}}^{[m]}u_{i,k_{+}'}^{[m]*}
    +\sum_{j\in\mathrm{B}}
     e^{i\bm{q}\cdot\bm{r}_{j}}
     t_{j,k_{+}}^{[m]}t_{j,k_{+}'}^{[m]*}
   \right),
   \nonumber
   \allowdisplaybreaks
   \\
   &
   N_{-+}^{(0)}(\bm{q};k_{-},k_{+}')
  \equiv
   \sum_{l=1}^{L}
   \tilde{N}_{-+}^{(0)}(l,\bm{q};k_{-},k_{+})
  =\frac{1}{\sqrt{L}}
   \left(
    -\sum_{i\in\mathrm{A}}
     e^{i\bm{q}\cdot\bm{r}_{i}}
     s_{i,k_{-}}^{[m]}u_{i,k_{+}'}^{[m]*}
    +\sum_{j\in\mathrm{B}}
     e^{i\bm{q}\cdot\bm{r}_{j}}
     v_{j,k_{-}}^{[m]}t_{j,k_{+}'}^{[m]*}
   \right).
   \label{E:deltaSqCoeff}
\end{align}
If we truncate the scattering operators as well at the order of $S^0$, the transverse and
longitudinal contributions to the dynamic structure factor,
involving a single and couple of magnons, respectively, are explicitly given by
\begin{align}
  &
   S^{\perp}(\bm{q};\omega)^{\![m]}
  =\frac{1}{2}\left[
     S^{+-}(\bm{q};\omega)^{\![m]}+S^{-+}(\bm{q};\omega)^{\![m]}
   \right]
  \allowdisplaybreaks
  \nonumber \\
  &\qquad\qquad\ \ \ 
  =\sum_{\sigma=\mp}
   \sum_{k_{\sigma}=1}^{L_{\sigma}}
   \left\{
    \frac{1}{2}
    \left|\varLambda_{\sigma}^{( \frac{1}{2})}(\bm{q};k_{\sigma})\right|^{2}
   +\mathrm{Re}
    \left[\varLambda_{\sigma}^{( \frac{1}{2})}(\bm{q};k_{\sigma})
          \varLambda_{\sigma}^{(-\frac{1}{2})}(\bm{q};k_{\sigma})^{*}\right]
   \right\}
   \delta\left(\hbar\omega-\varepsilon_{k_{\sigma}}^{\sigma[m]}\right),
   \label{E:SqwExpressionT}
   \allowdisplaybreaks
   \\
   &
   S^{\parallel}(\bm{q};\omega)^{\![m]}
  =\sum_{k_{-}=1}^{L_-}\sum_{k_{+}=1}^{L_+}
   \left|N_{-+}^{(0)}(\bm{q};k_{-},k_{+})\right|^{2}
   \delta\left(\hbar\omega-\varepsilon_{k_{-}}^{-[m]}-\varepsilon_{k_{+}}^{+[m]}\right),
   \label{E:SqwExpressionL}
\end{align}
where $\varLambda_\sigma^{(\pm\frac{1}{2})}$ read
$D_-^{(\pm\frac{1}{2})}$ ($\sigma=-$) or $C_+^{(\pm\frac{1}{2})}$ ($\sigma=+$).
\begin{figure}
\centering
\includegraphics[width=\linewidth]{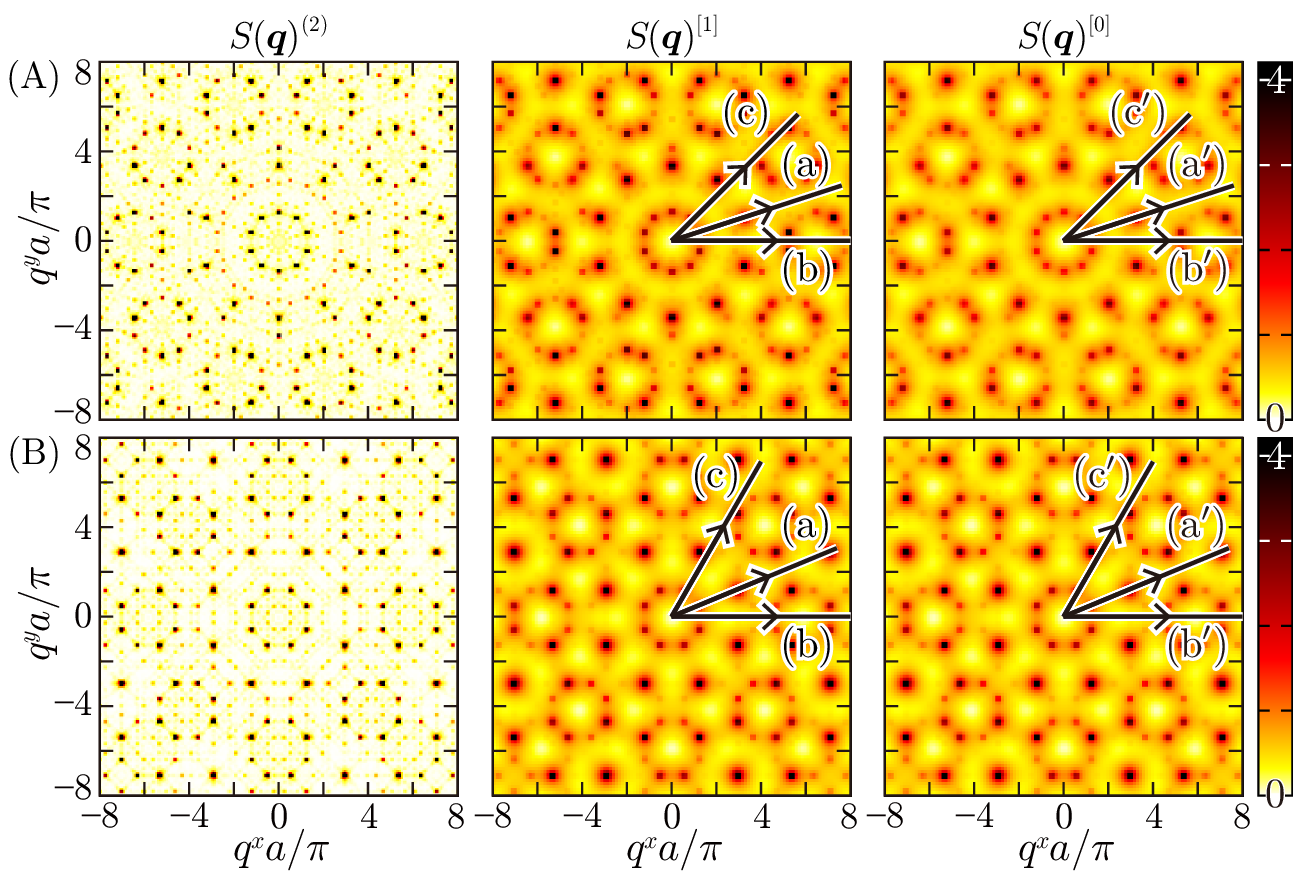}
\caption{%
         Contour plots of the static structure factors in
         the classical
         [$S(\bm{q})^{\!(2)}\equiv S^{zz}(\bm{q})^{\!(2)}$]
         and quantum
         [$S(\bm{q})^{\![m]}\equiv\sum_{\lambda=x,y,z}S^{\lambda\lambda}(\bm{q})^{\![m]}$]
         ground states, given by Eqs. \eqref{E:classicalSq} and \eqref{E:quantumSq},
         for the two-dimensional Penrose lattice of $L=601$ (A)
         and Ammann-Beenker lattice of $L=481$ (B).
         The paths (a) and (a${}^{\prime}$), (b) and (b${}^{\prime}$), and
         (c) and (c${}^{\prime}$) specified for $S(\bm{q})^{\![m]}$ denote the lines
         $q^{y}=\tan(\frac{\pi}{10}+\frac{2\pi}{10}n)q^{x}$,
         $q^{y}=\tan(               \frac{2\pi}{10}n)q^{x}$, and
         $q^{y}=\tan(\frac{\pi}{ 4}+\frac{2\pi}{10}n)q^{x}$
         $(n=0,1,2,\cdots)$, respectively, in (A), and
         $q^{y}=\tan(\frac{\pi}{ 8}+\frac{2\pi}{ 8}n)q^{x}$,
         $q^{y}=\tan(               \frac{2\pi}{ 8}n)q^{x}$, and
         $q^{y}=\tan(\frac{\pi}{ 3}+\frac{2\pi}{ 8}n)q^{x}$
         $(n=0,1,2,\cdots)$, respectively, in (B),
         each
         going along the strongest magnetic Bragg spots,
         avoiding every magnetic Bragg spot, and
         going along modest magnetic Bragg spots other than the strongest ones, respectively,
         in both (A) and (B).
         The thus-labeled paths in (A) and (B) correspond to those in
         Figures \ref{F:SqwPenrose} and \ref{F:SqwAB}, respectively.}
\label{F:Sq}
\end{figure}

   The static structure factors are available from the dynamic structure factors,
\begin{align}
   S^{\lambda\lambda}(\bm{q})^{\![m]}
  \equiv
   \frac{1}{L}
   \sum_{l,l'=1}^{L}
   e^{i\bm{q}\cdot(\bm{r}_{l}-\bm{r}_{l'})}
   {}^{\ [m]}\!\langle 0|
     \delta S_{\bm{r}_{l}}^{\lambda} \delta S_{\bm{r}_{l'}}^{\lambda}
   |0 \rangle^{\![m]}
  =
   \int_{-\infty}^{\infty}
   S^{\lambda\lambda}(\bm{q};\omega)^{\![m]}
   \hbar d\omega.
   \label{E:quantumSq}
\end{align}
For comparison, we define the classical static structure factors
\begin{align}
   S^{\lambda\lambda}(\bm{q})^{\!(2)}
  \equiv
   \frac{1}{L}
   \sum_{l,l'=1}^{L}
   e^{i\bm{q}\cdot(\bm{r}_{l}-\bm{r}_{l'})}
   {}^{\ (2)}\!\langle 0|
     S_{\bm{r}_{l}}^{\lambda}S_{\bm{r}_{l'}}^{\lambda}
   |0 \rangle^{\!(2)}
  =\left\{
    \begin{array}{ll}
     0 & (\lambda=x,y) \\
     \displaystyle
     \frac{1}{L}\sum_{l,l'=1}^{L}
     e^{i\bm{q}\cdot(\bm{r}_{l}-\bm{r}_{l'})}
     (-1)^{\sigma_{l}+\sigma_{l'}}
     S^{2} & (\lambda=z) \\
    \end{array}
   \right.
   \label{E:classicalSq}
\end{align}
as well,
where $\sigma_l$ takes $+1$ and $-1$ for $l\in\mathrm{A}$ and $l\in\mathrm{B}$, respectively.
The classical ground state $|0 \rangle^{\!(2)}$ has no spin fluctuation and therefore maximize
the longitudinal component.

\subsection{Static Structure Factors}
   Before investigating the dynamic structure factors as functions of momentum and energy,
we probe the static structure factors to reveal momenta corresponding to magnetic Bragg peaks.
We present those for the Penrose and Ammann-Beenker lattices in Figures \ref{F:Sq}(A) and
\ref{F:Sq}(B), respectively.
Demanding that the quantities \eqref{E:classicalSq} and \eqref{E:quantumSq} be real immediately
yields that
$S^{\lambda\lambda}(-\bm{q})^{\![m]}
=S^{\lambda\lambda}( \bm{q})^{\![m]}$
and
$S^{\lambda\lambda}(-\bm{q})^{\!(2)}
=S^{\lambda\lambda}( \bm{q})^{\!(2)}$.
An $n$-fold rotational symmetry of the background lattice ensures the same symmetry
in the momentum space,
$S^{\lambda\lambda}(C_n\bm{q})^{\![m]}
=S^{\lambda\lambda}(   \bm{q})^{\![m]}$
and
$S^{\lambda\lambda}(C_n\bm{q})^{\!(2)}
=S^{\lambda\lambda}(   \bm{q})^{\!(2)}$.
Therefore, an $n$-fold-rotation-invariant lattice yields a reciprocal lattice of
$\mathbf{C}_{n\mathrm{h}}$ point symmetry.
In two dimensions, $\mathbf{C}_{n\mathrm{h}}$ is isomorphic to
$\mathbf{C}_{2n}$ ($\mathbf{C}_{n}$) when $n$ is odd (even).
We are thus convinced of the ten- and eight-fold-rotation-invariant
$S(\bm{q})^{\!(2)}
\equiv
 \sum_{\lambda=x,y,z}S^{\lambda\lambda}(\bm{q})^{\!(2)}$
and
$S(\bm{q})^{\![m]}
\equiv
 \sum_{\lambda=x,y,z}S^{\lambda\lambda}(\bm{q})^{\![m]}$
for the Penrose [Figure \ref{F:Sq}(A)] and Ammann-Beenker [Figure \ref{F:Sq}(B)] lattices,
respectively.

   Quantum fluctuations dull all the magnetic Bragg peaks more or less, but their positions
remain unchanged, evidencing that a collinear long-range order of the $180^\circ$  N\'{e}el type
is robust on both Penrose and Ammann-Beenker lattices.
While the classical static structure factor $S(\bm{q})^{\!(2)}$ is made only of longitudinal
spin correlations, the quantum corrections $S(\bm{q})^{\![m]}$ are relevant to both
transverse and longitudinal components.
Both corrections are similarly effective on the momenta corresponding to the magnetic Bragg peaks
though the transverse components are dominant.
\begin{figure}
\centering
\includegraphics[width=\linewidth]{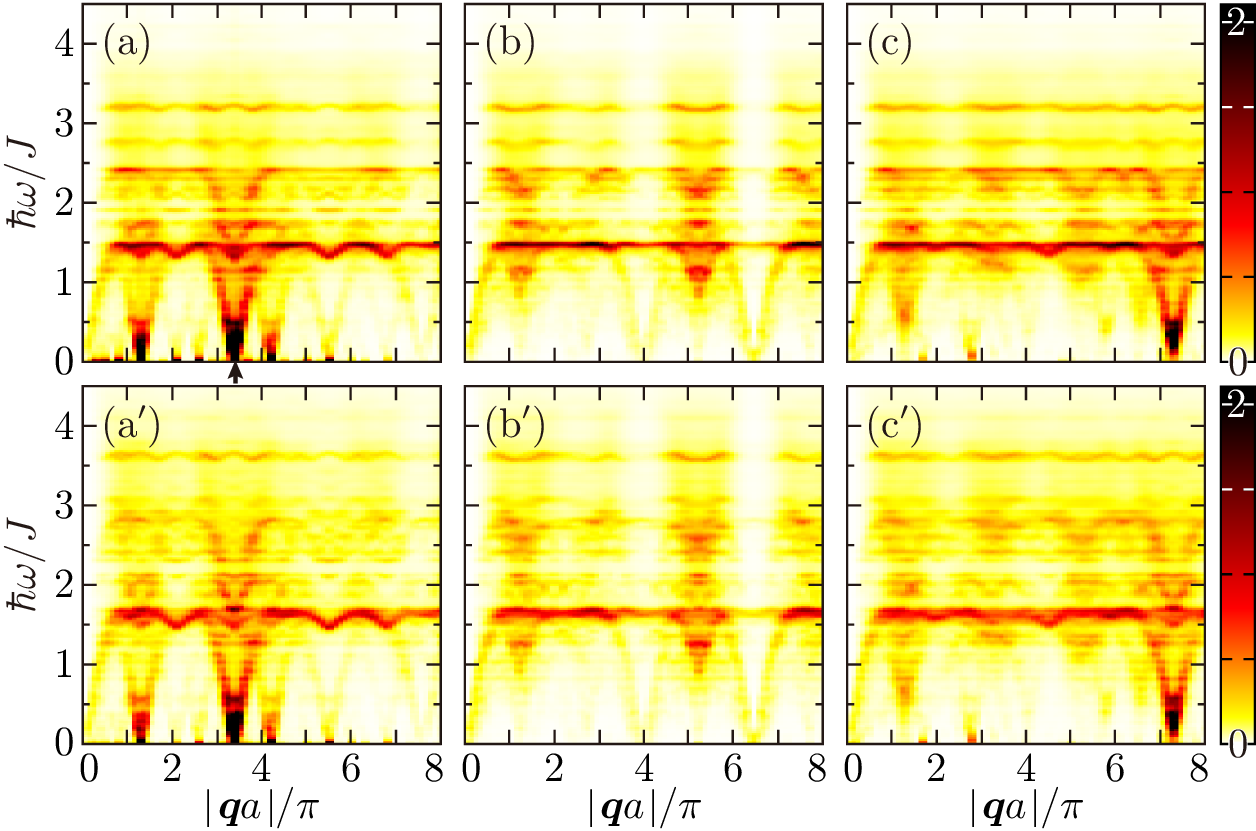}
\caption{%
         The dynamic structure factors
         $S(\bm{q};\omega)^{\![m]}
         \equiv
          S^\perp    (\bm{q};\omega)^{\![m]}
         +S^\parallel(\bm{q};\omega)^{\![m]}$
         in the LSW [$m=1$; (a), (b), and (c)] and 
                ISW [$m=0$; (a${}^{\prime}$), (b${}^{\prime}$), and (c${}^{\prime}$)]
         ground states, given by Eqs. \eqref{E:SqwExpressionT} and \eqref{E:SqwExpressionL},
         for the Penrose lattice of $L=11006$.
         The three paths in the momentum space,
         (a) and (a${}^{\prime}$), (b) and (b${}^{\prime}$), and (c) and (c${}^{\prime}$),
         are specified in Figure \ref{F:Sq}(A).
         Every spectral intensity is Lorentzian-broadened by a width of $0.01J$.
         The arrow indicates the momentum and energy in the panel (a) at which we make
         a perpendicular-space analysis of the symmetrized site-resolved dynamic structure
         factor \eqref{E:siteresolvedSqwRsymmetrized} in
         Figure \ref{F:perpCP_LSWgoldstone}.}
\label{F:SqwPenrose}
\end{figure}
\begin{figure}
\centering
\includegraphics[width=\linewidth]{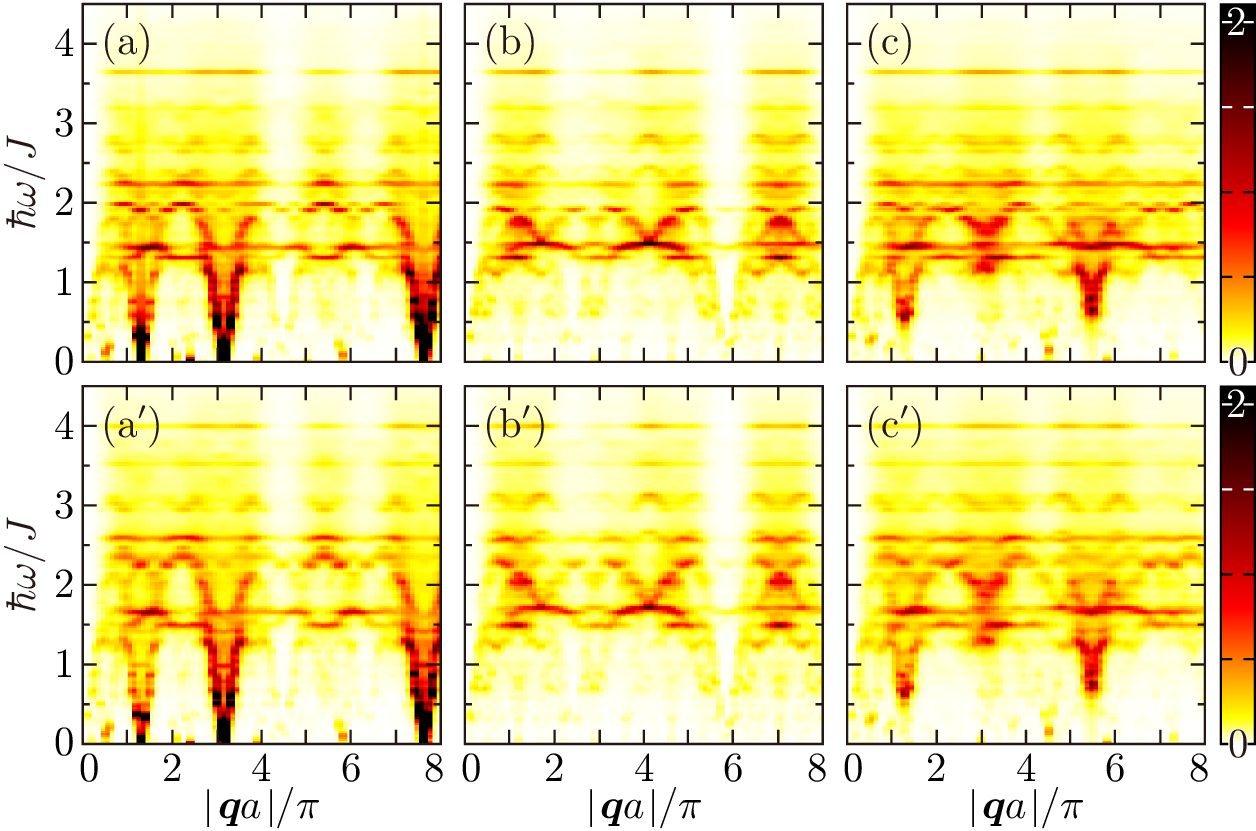}
\caption{%
         The dynamic structure factors
         $S(\bm{q};\omega)^{\![m]}
         \equiv
          S^\perp    (\bm{q};\omega)^{\![m]}
         +S^\parallel(\bm{q};\omega)^{\![m]}$
         in the LSW [$m=1$; (a), (b), and (c)] and 
                ISW [$m=0$; (a${}^{\prime}$), (b${}^{\prime}$), and (c${}^{\prime}$)]
         ground states, given by Eqs. \eqref{E:SqwExpressionT} and \eqref{E:SqwExpressionL},
         for the Ammann-Beenker lattice of $L=10457$.
         The three paths in the momentum space,
         (a) and (a${}^{\prime}$), (b) and (b${}^{\prime}$), and (c) and (c${}^{\prime}$),
         are specified in Figure \ref{F:Sq}(B).
         Every spectral intensity is Lorentzian-broadened by a width of $0.01J$.}
\label{F:SqwAB}
\end{figure}

\subsection{Dynamic Structure Factors}
   Figures \ref{F:SqwPenrose} and \ref{F:SqwAB} show SW calculations of
the dynamic structure factors
$S(\bm{q};\omega)^{\![m]}
\equiv
 S^\perp    (\bm{q};\omega)^{\![m]}
+S^\parallel(\bm{q};\omega)^{\![m]}$
for the $S=\frac{1}{2}$ Heisenberg antiferromagnets \eqref{E:Ham} on the Penrose and
Ammann-Beenker lattices, respectively,
within and beyond the harmonic-oscillator approximation.
The transverse component $S^\perp(\bm{q};\omega)^{\![m]}$ predominates
the longitudinal component $S^\parallel(\bm{q};\omega)^{\![m]}$ in spectral weight,
especially for the Penrose lattice, when compared to the Ammann-Beenker lattice, and
relatively at $m=0$, when compared to $m=1$ [See Appendix].
For both lattices, a linear soft mode appears at every magnetic Bragg wavevector and
nearly or fairly flat bands, signifying magnetic excitations localized in some way,
lie at several different energies in a self-similar manner.
There is a possibility of a gapless mode and a flat mode coexisting in periodic systems as well,
as is the case with ordered kagome-lattice antiferromagnets \cite{M064428,C144415,Y065004}.
Schwinger bosons \cite{M064428} and Holstein-Primakoff bosons \cite{C144415,Y065004} applied to
nearest-neighbor Heisenberg antiferromagnets on the regular kagome lattice both yield
a Nambu-Goldstone mode of the long-range N\'eel order, whether of the $\sqrt{3}\times\sqrt{3}$ or
$\bm{Q}=\bm{0}$ type, and a dispersionless branch consisting of excitations localized to
an arbitrary hexagon of nearest-neighbor spins \cite{H2899}.
The wholly flat band is gapless or gapfull according as whether or not SWs are brought into
interaction and/or modification \cite{Y065004}.
However, there is no self-similar structure in these findings.
\begin{figure}
\centering
\includegraphics[width=\linewidth]{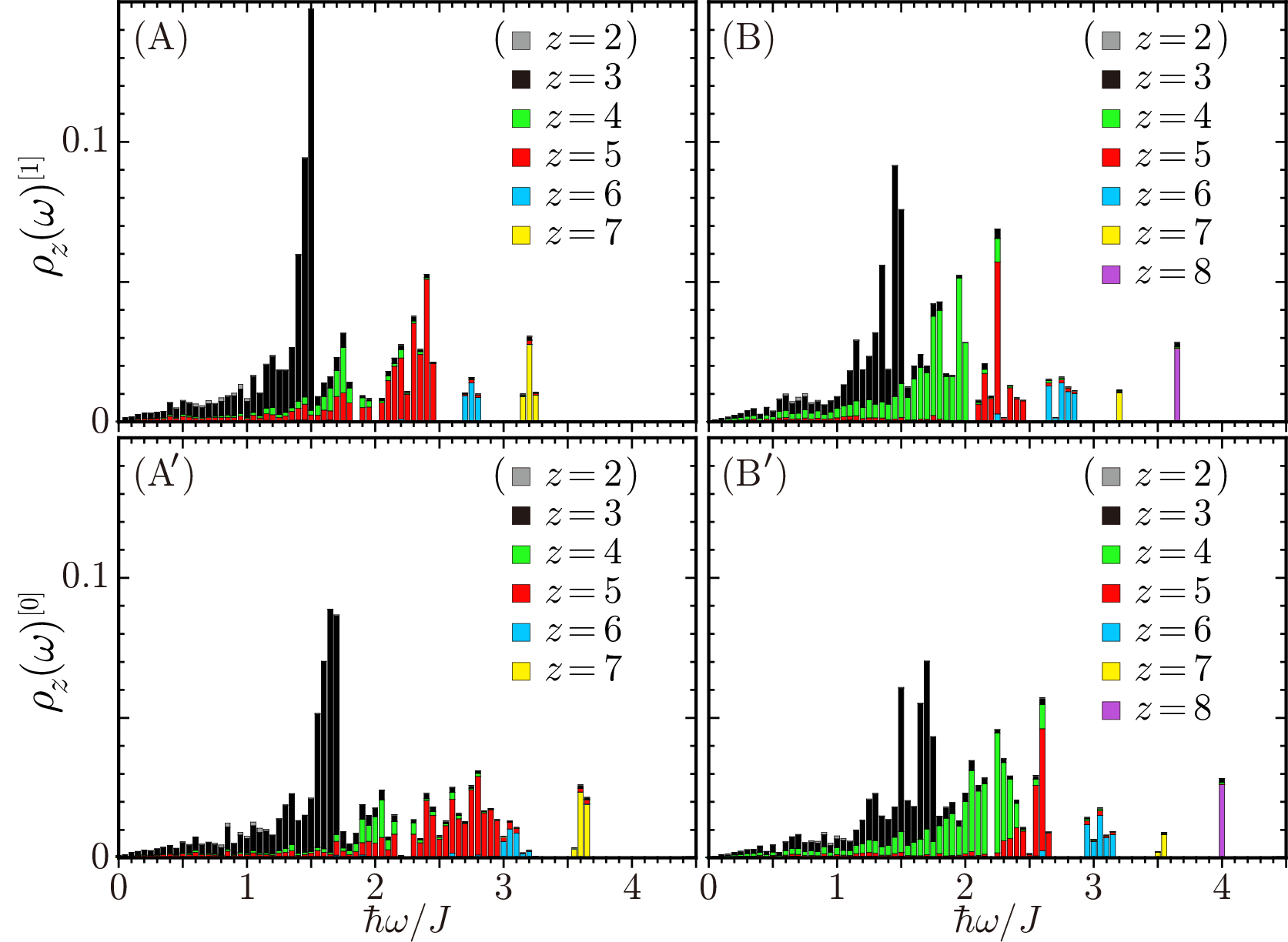}
\caption{%
         The site-resolved density of states
         $\sum_{z}\rho_{z}(\omega)^{[m]}$
         of the LSW [$m=1$; (A) and (B)] and ISW [$m=0$; (A${}^{\prime}$) and (B${}^{\prime}$)]
         excitation spectra, given by Eq. \eqref{E:SWDOS}, for the
         Penrose lattice of $L=11006$ [(A) and (A${}^{\prime}$)] and
         Ammann-Beenker lattice of $L=10457$ [(B) and (B${}^{\prime}$)].
         $z$ should take $3$ to $7$ and $3$ to $8$ on the Penrose and Ammann-Beenker lattices,
         respectively.
         Note that bicoordinated vertices are artifacts on finite clusters under the open
         boundary condition.}
\label{F:HeisenbergDOS}
\end{figure}

   When we compare the scattering bands of dispersionless character for the Penrose and
Ammann-Beenker lattices, both lying around $\hbar\omega=1.5J$ in the LSW formalism and
shifting upward in the ISW formalism, those for the Penrose lattice are
much flatter in every direction and much stronger in spectral weight.
In order to reveal to what types of magnetic excitations the Penrose lattice owe its nearly flat
band and the Ammann-Beenker lattice owe its similar but different bandlike segments of modulated
spectral weight, we plot in Figure \ref{F:HeisenbergDOS} the site-resolved density of states
\begin{align}
   &\!\!
   \rho(\omega)^{\![m]}
  =\sum_{l=1}^{L}
   \rho_{l}(\omega)^{\![m]}
  =\sum_{z=z_{\mathrm{min}}}^{z_{\mathrm{max}}}
   \rho_{z}(\omega)^{\![m]}
  =\sum_{z=z_{\mathrm{min}}}^{z_{\mathrm{max}}}
   \frac{1}{L}
   \allowdisplaybreaks
   \nonumber \\
   &\!\!
   \times
   \left\{
     \sum_{k_{-}=1}^{L_{-}}
     \frac{ \sum_{i(z_i=z)}\left|s_{i,k_{-}}^{[m]}\right|^{2}
           +\sum_{j(z_j=z)}\left|v_{j,k_{-}}^{[m]}\right|^{2}}
          { \sum_{i\in \mathrm{A}}\left|s_{i,k_{-}}^{[m]}\right|^{2}
           +\sum_{j\in \mathrm{B}}\left|v_{j,k_{-}}^{[m]}\right|^{2}}
     \delta\left(\hbar\omega-\varepsilon_{k_{-}}^{-[m]}\right)
    +\sum_{k_{+}=1}^{L_{+}}
     \frac{ \sum_{i(z_i=z)}\left|u_{i,k_{+}}^{[m]}\right|^{2}
           +\sum_{j(z_j=z)}\left|t_{j,k_{+}}^{[m]}\right|^{2}}
          { \sum_{i\in \mathrm{A}}\left|u_{i,k_{+}}^{[m]}\right|^{2}
           +\sum_{j\in \mathrm{B}}\left|t_{j,k_{+}}^{[m]}\right|^{2}}
     \delta\left(\hbar\omega-\varepsilon_{k_{+}}^{+[m]}\right)
   \right\}
   \label{E:SWDOS}
\end{align}
of their LSW ($m=1$) and ISW ($m=0$) excitation spectra.
For the Ising antiferromagnet, any spin deviation is immobile and its SW excitation spectrum
exactly consists of $z_{\mathrm{max}}-z_{\mathrm{min}}+1$ discrete eigenlevels of energy $zSJ$.
With this in mind, we take a keen interest in the highly degenerate LSW eigenstates of energy
$\hbar\omega=1.5J$ for the Heisenberg antiferromagnet on the Penrose lattice
[Figure \ref{F:HeisenbergDOS}(A)],
which are essentially composed of tricoordinated sites only and therefore whose energy can be
interpreted as $zSJ=\frac{3}{2}J$.
Hence it follows that they are no different from immobile antiferromagnons emergent on
tricoordinated sites in the Ising limit.
Then the lowest-lying nearly flat scattering band of $\hbar\omega=1.5J$ detected in
Figures \ref{F:SqwPenrose}(a)--\ref{F:SqwPenrose}(c) reads as an evidence of antiferromagnons
strongly confined within sites of $z=3$.
For the Heisenberg antiferromagnet on the Ammann-Beenker lattice, on the other hand,
much less states concentrate on a particular energy.
Indeed there are still LSW eigenstates of energy $\hbar\omega=1.5J$ existing, but they are here
composed of quadricoordinated as well as tricoordinated sites, implying weaker magnon confinement.
When LSWs are brought into interaction, the eigenspectrum and therefore scattering spectrum
generally shift upward in energy.
\begin{figure}
\centering
\includegraphics[width=0.7\linewidth]{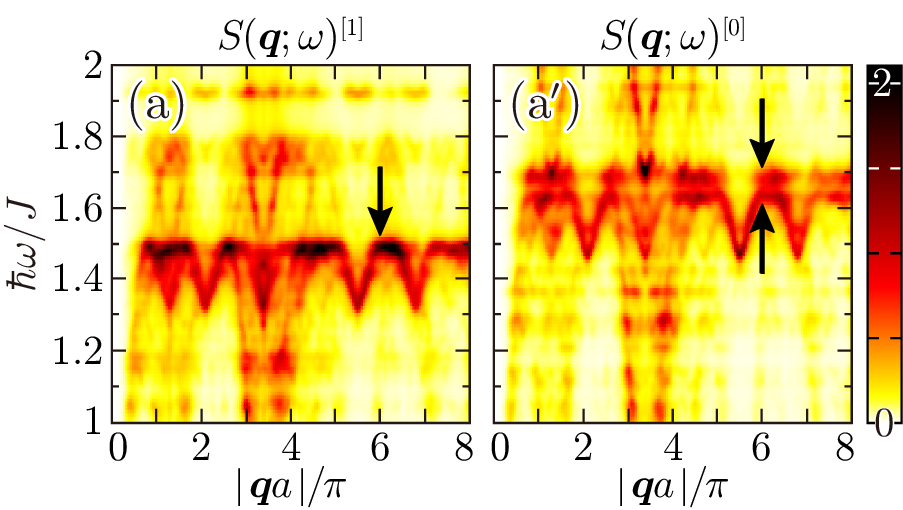}
\caption{%
         Particular-energy enlarged views of Figures \ref{F:SqwPenrose}(a) and
         \ref{F:SqwPenrose}(a${}^{\prime}$).
         The arrows in the panels (a) and (a${}^{\prime}$) indicate the momenta and energies
         at which we make perpendicular-space analyses of the symmetrized site-resolved dynamic
         structure factor \eqref{E:siteresolvedSqwRsymmetrized} in
         Figures \ref{F:perpCP_LSWflat} and \ref{F:perpCP_ISWflats}.}
\label{F:SqwLSWvsISW}
\end{figure}

   The lowest-lying nearly flat scattering band is thus characteristic of the Penrose lattice.
It signifies that there exist intriguing antiferromagnons strongly confined within tricoordinated
sites.
In order to investigate their properties in more detail, we enlarge
Figures \ref{F:SqwPenrose}(a) and \ref{F:SqwPenrose}(a${}^{\prime}$) around the particular
energy $\hbar\omega=\frac{3}{2}J$ in Figures \ref{F:SqwLSWvsISW}(a) and
\ref{F:SqwLSWvsISW}(a${}^{\prime}$), respectively.
$O(S^0)$ quantum fluctuations not only push up in energy the nearly flat band but also split it in
two, lying at about $1.63J$ and $1.70J$.
In Figure \ref{F:HeisenbergDOS}(A${}^{\prime}$), the ISW eigenstates of $z=3$ character no longer
concentrate on a particular energy as the LSW ones, indeed, and their density exhibits
a relatively loose peak at around $1.65J$.
Such an upward quantum renormalization of the LSW excitation energy is usual with
collinear antiferromagnets on bipartite lattices.
If we consider the same Hamiltonian as \eqref{E:Ham} on two-dimensional bipartite periodic lattices
such as the quadricoordinated square lattice and tricoordinated honeycomb lattice
whose two sublattices are equivalent to each other, $L_{\mathrm{A}}=L_{\mathrm{B}}=L/2\equiv N$,
whether $L$ is infinite or finite, its up-to-$O(S^0)$ bilinear ISW Hamiltonian in the form of
\eqref{E:HBL[0]} is diagonalizable as \cite{N034714}
\begin{align}
   &
   \mathcal{H}_{\mathrm{BL}}^{[0]}
 =-NzJ(S-\epsilon)^2
  +zJ(S+\epsilon)
   \sum_{\nu=1}^N
   \sqrt{1-|\gamma_{\bm{k}_\nu}|^2}
   \sum_{\sigma=\mp}
   \alpha_{\bm{k}_\nu}^{\sigma\dagger}\alpha_{\bm{k}_\nu}^\sigma;
   \allowdisplaybreaks
   \nonumber \\
   &
   \epsilon
  \equiv
   \frac{1}{2}
  -\frac{1}{2N}
   \sum_{\nu=1}^N
   \sqrt{1-|\gamma_{\bm{k}_\nu}|^2},\ 
   \gamma_{\bm{k}_\nu}
  \equiv
   \frac{1}{z}
   \sum_{\kappa=1}^z
   e^{i\bm{k}_\nu\cdot\bm{\delta}(\kappa)},
   \label{E:epsilon&gamma_k}
\end{align}
with the set of vectors pointing to $z$ nearest-neighbor sites to $\bm{r}_l$,
$\bm{\delta}(1)$ to $\bm{\delta}(z)$, being no longer dependent on the site index $l$,
where the degenerate ferromagnetic ($\sigma=-$) and antiferromagnetic ($\sigma=+$) quasiparticle
magnons are made of the Holstein-Primakoff bosons \eqref{E:HPboson} as
\begin{align}
   \left\{
   \begin{array}{l}
    \displaystyle
    \alpha_{\bm{k}_\nu}^+
   =a_{\bm{k}_\nu}^\dagger\frac{\gamma_{\bm{k}_\nu}}{|\gamma_{\bm{k}_\nu}|}
                          \mathrm{sinh}\theta_{\bm{k}_\nu}
   +b_{\bm{k}_\nu}        \mathrm{cosh}\theta_{\bm{k}_\nu} \\
    \displaystyle
    \alpha_{\bm{k}_\nu}^-
   =a_{\bm{k}_\nu}        \mathrm{cosh}\theta_{\bm{k}_\nu}
   +b_{\bm{k}_\nu}^\dagger\frac{\gamma_{\bm{k}_\nu}}{|\gamma_{\bm{k}_\nu}|}
                          \mathrm{sinh}\theta_{\bm{k}_\nu} \\
   \end{array}
   \right.,\ 
   \left\{
   \begin{array}{l}
    \displaystyle
    a_{\bm{k}_\nu}^\dagger
   =\frac{1}{\sqrt{N}}\sum_{i\in\mathrm{A}}
    \mathrm{e}^{i\bm{k}_\nu\cdot\bm{r}_i}a_i^\dagger \\
    \displaystyle
    b_{\bm{k}_\nu}
   =\frac{1}{\sqrt{N}}\sum_{j\in\mathrm{B}}
    \mathrm{e}^{\mathrm{i}\bm{k}_\nu\cdot\bm{r}_j}b_j \\
   \end{array}
   \right.,\ 
   \left\{
   \begin{array}{l}
    \displaystyle
    \mathrm{cosh}2\theta_{\bm{k}_\nu}
   =\frac{1}{\sqrt{1-|\gamma_{\bm{k}_\nu}|^2}} \\ 
    \displaystyle
    \mathrm{sinh}2\theta_{\bm{k}_\nu}
   =\frac{|\gamma_{\bm{k}_\nu}|}{\sqrt{1-|\gamma_{\bm{k}_\nu}|^2}} \\
   \end{array}
   \right.,
   \label{E:BT}
\end{align}
whether within or beyond the harmonic-oscillator approximation.
Interestingly, besides sharing the Bogoliubov transformation, the LSWs and up-to-$O(S^0)$ ISWs
have similar energies
$\varepsilon_{\bm{k}_\nu}^{[1]}=z S          \sqrt{1-|\gamma_{\bm{k}_\nu}|^2}J$ and
$\varepsilon_{\bm{k}_\nu}^{[0]}=z(S+\epsilon)\sqrt{1-|\gamma_{\bm{k}_\nu}|^2}J$
\cite{Y094412,S216003}, respectively,
which are obtainable from each other via a wavevector-independent renormalization.
For the periodic square-lattice antiferromagnet, for instance,
the set of wavevectors as good quantum numbers, $\bm{k}_1$ to $\bm{k}_N$, can be expressed as
\begin{align}
   \bm{k}_\nu
  \equiv
   \bm{k}_{(\nu^{(1)},\nu^{(2)})}
  =\sum_{n=1}^2
   \left(
    \frac{\nu^{(n)}}{\sqrt{N}}-\frac{1}{2}
   \right)
   \tilde{\bm{g}}(n);\ 
   \nu
  =\sum_{n=1}^2
   \bigl(
    \nu^{(n)}-1
   \bigr)
   \bigl(
    \sqrt{N}
   \bigr)^{n-1}
  +1,\ 
   \nu^{(n)}
  =1,2,\cdots,\sqrt{N}
   \label{E:vectork(nu)}
\end{align}
in terms of the primitive translation vectors in the reciprocal space corresponding to
each sublattice A or B,
\begin{align}
   \tilde{\bm{g}}(1)
  =\frac{\pi}{a}
   \frac{\tilde{\bm{\delta}}(1)}{a},\ 
   \tilde{\bm{g}}(2)
  =\frac{\pi}{a}
   \frac{\tilde{\bm{\delta}}(2)}{a};\ 
   \tilde{\bm{\delta}}(1)
  \equiv
   \bm{\delta}(1)+\bm{\delta}(2),\ 
   \tilde{\bm{\delta}}(2)
  \equiv
   \bm{\delta}(1)-\bm{\delta}(2);\ 
   \tilde{\bm{\delta}}(n)\cdot\tilde{\bm{g}}(n')
  =2\pi\delta_{nn'},
   \label{E:PTVinHalfKspace}
\end{align}
and its up-to-$O(S^0)$ antiferromagnons exhibit no dispersion on the magnetic Brillouin zone
boundary, i.e., along the line connecting
$\frac{\pi}{a}(1,0)$ and $\frac{\pi}{a}(\frac{1}{2},\frac{1}{2})$.
In the case of $S=\frac{1}{2}$, the $O(S^0)$ quantum renormalization factor
$\varepsilon_{\bm{k}_\nu}^{[0]}/\varepsilon_{\bm{k}_\nu}^{[1]}$ reads $1.157947$.
The up-to-$O(S^0)$ modified SW theory \cite{Y094412} and equivalent Auerbach-Arovas'
Schwinger-boson mean-field theory \cite{A617} consequently derive such quantum renormalization
of SW ridges in the dynamic structure factor.
If we discuss the SW dynamics in the Penrose-lattice Heisenberg antiferromagnet by analogy,
its $O(S^0)$ quantum renormalization factor
$\varepsilon_{\bm{k}_\nu}^{[0]}/\varepsilon_{\bm{k}_\nu}^{[1]}$
is indeed larger than unity but looks blurry, extending between
$1.63/1.5\simeq 1.09$ and $1.70/1.5\simeq 1.13$.
In order to find out why such a split is generated in the lowest-lying nearly flat scattering band
and clarify whether or not the split two are still confined within tricoordinated sites,
we move to the perpendicular space.

\subsection{Perpendicular-Space Analysis}
   Suppose we define the site-resolved dynamic structure factors
\begin{align}
   &
   S^{\lambda\lambda}(\bm{q};\omega)^{\![m]}
  \equiv
   \sum_{l=1}^L
   S^{\lambda\lambda}(\bm{q};\omega)^{\![m]}\bigl|_{l}
  =\sum_{l,l'=1}^{L}
   \frac{e^{i\bm{q}\cdot(\bm{r}_{l}-\bm{r}_{l'})}}{2\pi\hbar L}
   \int_{-\infty}^{\infty}
   {}^{\ [m]}\!\langle 0|
     \delta S_{\bm{r}_{l}}^{\lambda}(t) \delta S_{\bm{r}_{l'}}^{\lambda}
   |0 \rangle^{\![m]}
   e^{i\omega t}dt.
   \label{E:siteresolvedSqw}
\end{align}
They can be expressed in terms of the coefficients explicitly given in \eqref{E:deltaSqCoeff} as
\begin{align}
   &
   S^{\perp}(\bm{q};\omega)^{\![m]}\bigl|_{l}
  \equiv
   S^{xx}(\bm{q};\omega)^{\![m]}\bigl|_{l}+S^{yy}(\bm{q};\omega)^{\![m]}\bigl|_{l}
  =\frac{1}{2}
   \sum_{\sigma=\mp}\sum_{k_{\sigma}=1}^{L_{\sigma}}
    \left[
     \tilde{\varLambda}_{\sigma}^{(\frac{1}{2})}(l,\bm{q};k_{\sigma})
     \varLambda_{\sigma}^{(\frac{1}{2})}(\bm{q};k_{\sigma})^{*}
    \right.
   \allowdisplaybreaks
   \nonumber \\
   &\qquad\qquad\quad
   +\left.
      \tilde{\varLambda}_{\sigma}^{(\frac{1}{2})}(l,\bm{q};k_{\sigma})
      \varLambda_{\sigma}^{(-\frac{1}{2})}(\bm{q};k_{\sigma})^{*}
     +\varLambda_{\sigma}^{(\frac{1}{2})}(\bm{q};k_{\sigma})
      \tilde{\varLambda}_{\sigma}^{(-\frac{1}{2})}(l,\bm{q};k_{\sigma})^{*}
    \right]
   \delta\left(\hbar\omega-\varepsilon_{k_{\sigma}}^{\sigma[m]}\right),
   \label{E:siteresolvedSqwT}
   \allowdisplaybreaks
   \\
   &
   S^{\parallel}(\bm{q};\omega)^{\![m]}\bigl|_{l}
  \equiv
   S^{zz}(\bm{q};\omega)^{\![m]}\bigl|_{l}
  =\sum_{k_{-}=1}^{L_{-}}\sum_{k_{+}=1}^{L_{+}}
   \left[
    \tilde{N}_{-+}^{(0)}(l,\bm{q};k_{-},k_{+})
    N_{-+}^{(0)}(\bm{q};k_{-},k_{+})^{*}
   \right]
   \delta\left(\hbar\omega-\varepsilon_{k_{-}}^{-[m]}-\varepsilon_{k_{+}}^{+[m]}\right),
   \label{E:siteresolvedSqwL}
\end{align}
where $\tilde{\varLambda}_\sigma^{(\pm\frac{1}{2})}$ read
$\tilde{D}_-^{(\pm\frac{1}{2})}$ ($\sigma=-$) or $\tilde{C}_+^{(\pm\frac{1}{2})}$ ($\sigma=+$)
just like $\varLambda_{\sigma}^{(\pm\frac{1}{2})}$ in \eqref{E:SqwExpressionT}.
In order to recover the point symmetry $\mathbf{P}$ of the original dynamic structure factor
$
   S(\bm{q};\omega)^{\![m]}
  \equiv
   \sum_{\lambda=x,y,z}S^{\lambda\lambda}(\bm{q};\omega)^{\![m]}
$
and thereby extract site-resolved spectral weights of real definite from it,
we resymmetrize the site-resolved dynamic structure factors
$
   S(\bm{q};\omega)^{\![m]}\bigl|_{l}
  \equiv
   \sum_{\lambda=x,y,z}S^{\lambda\lambda}(\bm{q};\omega)^{\![m]}\bigl|_{l}
$ 
into
\begin{align}
   \overline{S}(\bm{q};\omega)^{\![m]}\bigl|_{l}
  \equiv
   \frac{1}{g^{\mathbf{P}}}\sum_{p\in\mathbf{P}}
   S(p\bm{q};\omega)^{\![m]}\bigl|_{l},
   \label{E:siteresolvedSqwRsymmetrized}
\end{align}
where $g^{\mathbf{P}}$ is the order of $\mathbf{P}$.
The Penrose lattice demands that $\mathbf{P}=\mathbf{C}_{5\mathrm{h}}$.
\begin{figure}[h]
\centering
\includegraphics[width=\linewidth]{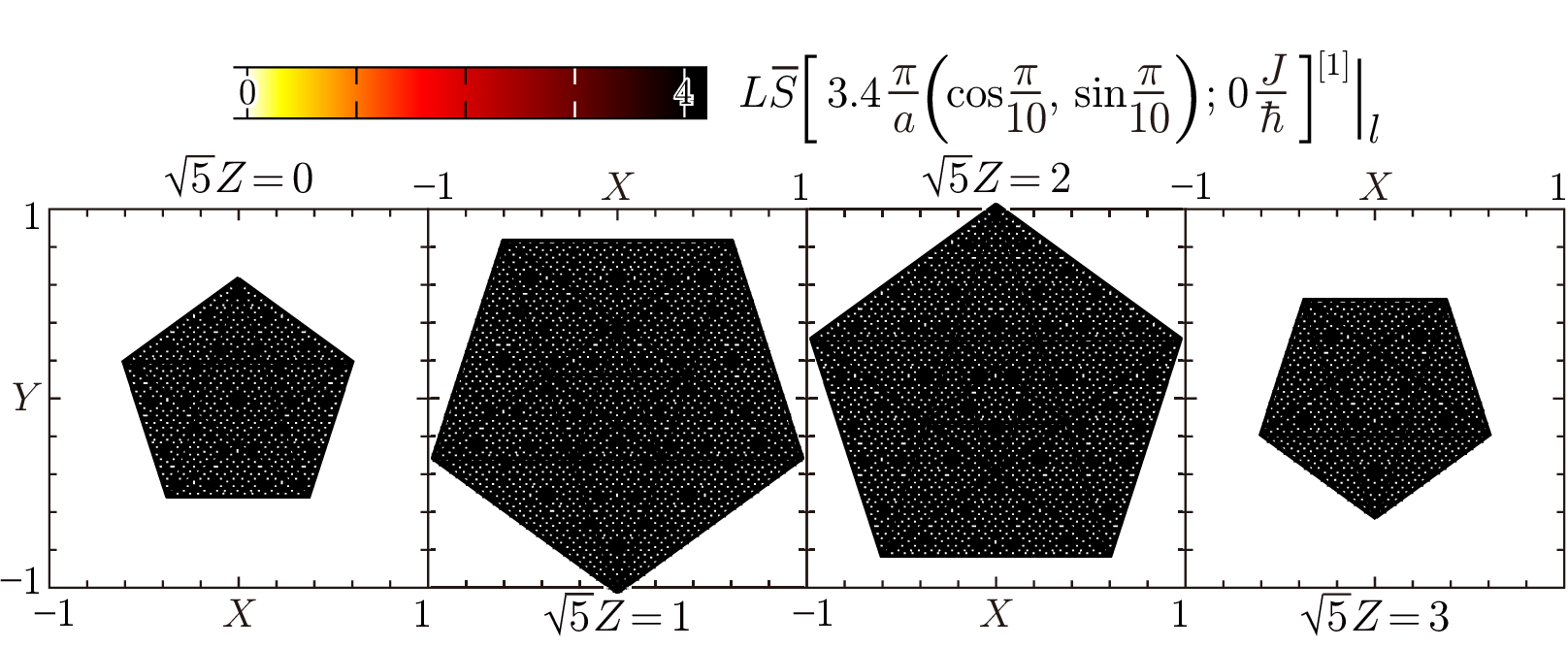}
\caption{%
         Contour plots of the symmetrized site-resolved dynamic structure factor
         $\overline{S}(\bm{q};\omega)^{\![1]}\bigl|_{l}$
         \eqref{E:siteresolvedSqwRsymmetrized} multiplied by the number of sites
         at $(q^{x},q^{y})=3.4\frac{\pi}{a}(\cos\frac{\pi}{10},\sin\frac{\pi}{10})$ 
         and $\hbar\omega=0$
         for Figure \ref{F:SqwPenrose}(a).}
\label{F:perpCP_LSWgoldstone}
\end{figure}

   In the context of making a perpendicular-space analysis of the flat bands signifying
confined states, it is worth observing linear soft modes similarly.
Figure \ref{F:perpCP_LSWgoldstone} is a perpendicular-space mapping of
$L\overline{S}(\bm{q};\omega)^{\![1]}\bigl|_{l}$ at momenta corresponding to the strongest
magnetic Bragg spots.
The spectral weighting is uniform and over the entire perpendicular space, evidencing that
soft modes at low energies involve all sites of $z=3$ to $7$.
The plot of $L\overline{S}(\bm{q};\omega)^{\![0]}\bigl|_{l}$ at the same momenta and zero energy
remains almost the same.
\begin{figure}
\centering
\includegraphics[width=\linewidth]{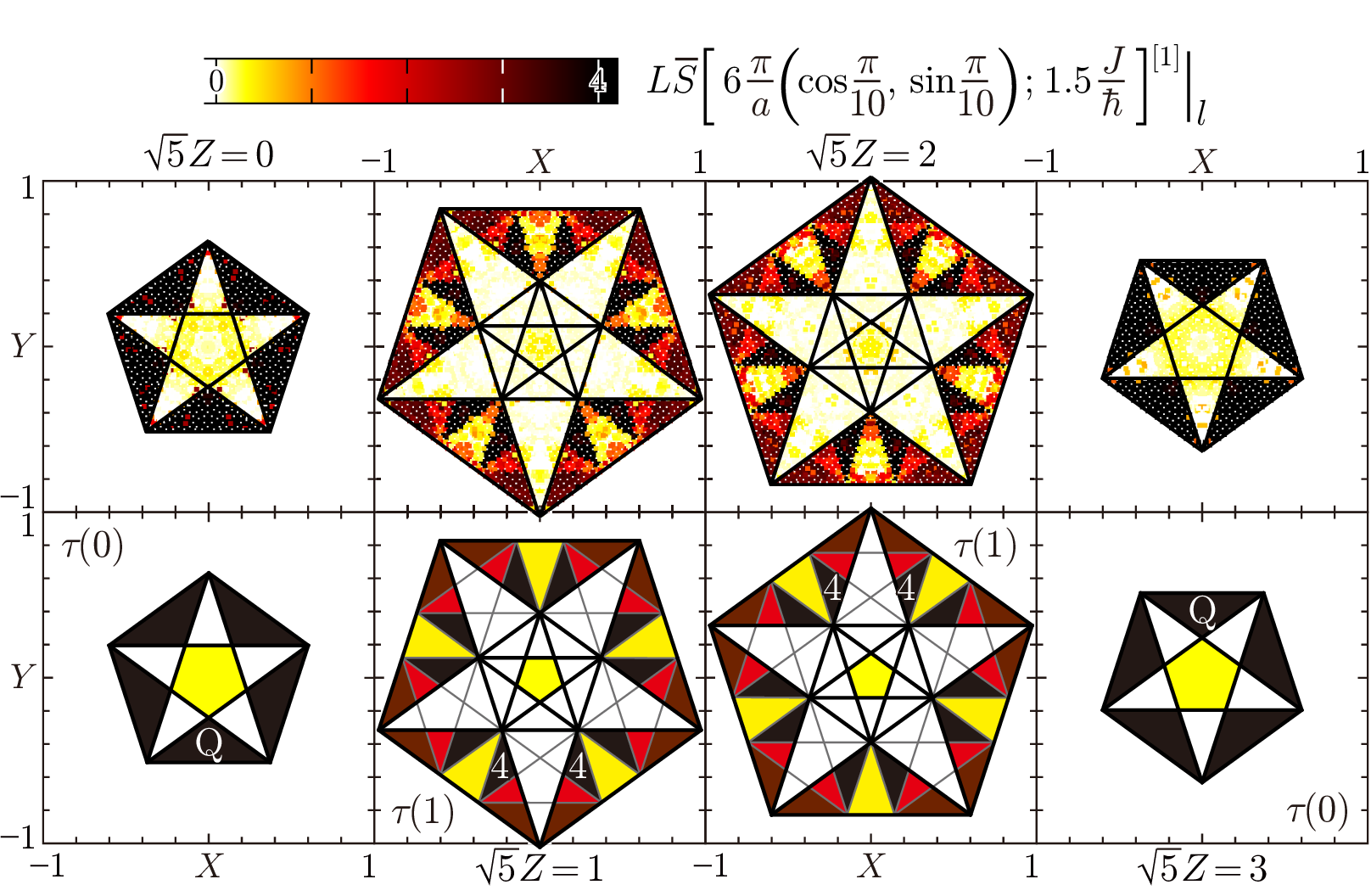}
\caption{%
         Contour plots of the symmetrized site-resolved dynamic structure factor
         $\overline{S}(\bm{q};\omega)^{\![1]}\bigl|_{l}$
         \eqref{E:siteresolvedSqwRsymmetrized} multiplied by the number of sites
         at $(q^{x},q^{y})=6\frac{\pi}{a}(\cos\frac{\pi}{10},\sin\frac{\pi}{10})$ and
         $\hbar\omega=1.5J$ for Figure \ref{F:SqwLSWvsISW}(a).
         LSW findings (the upper four panels) are compared with the $8$ types of domains
         $\tau(0)=\mathrm{D}$ to S3 ($\sqrt{5}Z=0,3$) or the $15$ types of subdomains
         $\tau(1)=1$ to $15$ ($\sqrt{5}Z=1,2$) illustrated with Figure \ref{F:Penrose}(d)
         (the lower four panels).}
\label{F:perpCP_LSWflat}
\end{figure}

   With the above in mind, let us observe the lowest-lying nearly flat scattering band peculiar
to the Penrose lattice in its perpendicular space.
Figure \ref{F:perpCP_LSWflat} is a perpendicular-space mapping of
$L\overline{S}(\bm{q};\omega)^{\![1]}\bigl|_{l}$ at
the wavevector $(q^{x},q^{y})=6\frac{\pi}{a}(\cos\frac{\pi}{10},\sin\frac{\pi}{10})$ and
energy $\hbar\omega=1.5J$ on the nearly flat LSW-scattering band [\ref{F:SqwLSWvsISW}(a)].
Now the spectral weighting is far from uniform and localizes on the D and Q domains in
the perpendicular space labeled by the $\tau(0)$ indices, evidencing that
the LSW nearly flat mode at around the particular energy $\frac{3}{2}J$ belongs to
tricoordinated sites.
A careful observation leads us to find out that the D domains are painted with several different
colors compatibly with sublabels $\tau(1)$ as well.
In terms of the $\tau(1)$ language, the spectral weights of the lowest-lying nearly flat LSW
scattering band center on the subdomains $4$, $5$, and $6$,
the former one of which resides in the $\tau(0)=\mathrm{D}$ domain,
while the rest two of which reside in the $\tau(0)=\mathrm{Q}$ domain,
and much less ooze out to the subdomains $1$ to $3$,
all residing in the $\tau(0)=\mathrm{D}$ domain.
If we sum up the coordination numbers of nearest-neighbor ($R=1$) sites at $\bm{r}_{l'}$
to its core ($R=0$) site at $\bm{r}_{l}$,
the three subdomains $4$ to $6$ in the $\tau(1)$ language can be classified into two groups,
$\{5\}$ with $\sum_{l'}z_{l'}(1)=16$ and $\{4,6\}$ with $\sum_{l'}z_{l'}(1)=17$
(See Table \ref{T:vertexR1} again).
This categorization plays a key role in understanding the split in the nearly flat scattering band
on the occurrence of interaction between LSWs.
\begin{figure}
\centering
\includegraphics[width=\linewidth]{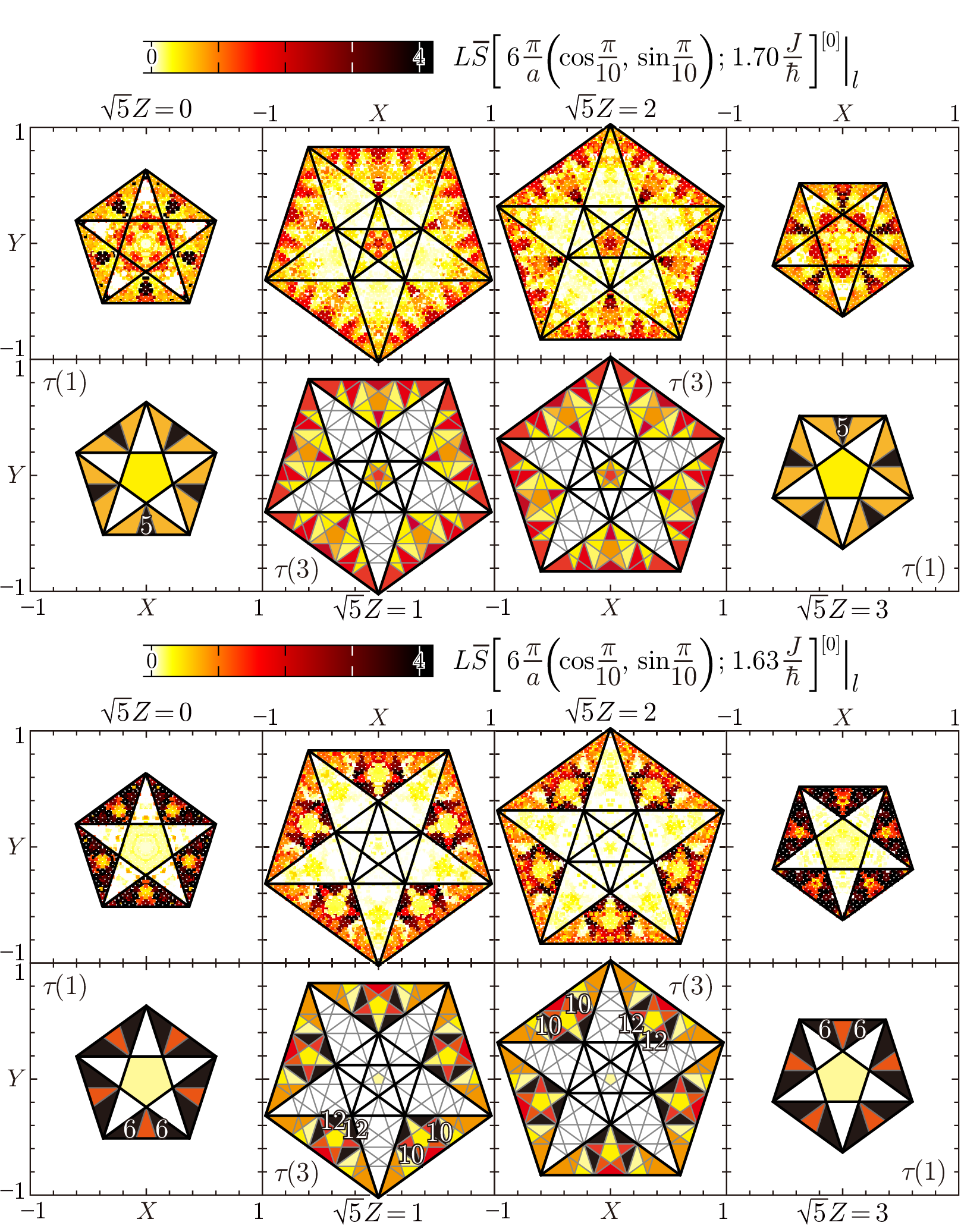}
\caption{%
         Contour plots of the symmetrized site-resolved dynamic structure factor
         $\overline{S}(\bm{q};\omega)^{\![0]}\bigl|_{l}$
         \eqref{E:siteresolvedSqwRsymmetrized} multiplied by the number of sites
         at $(q^{x},q^{y})=6\frac{\pi}{a}(\cos\frac{\pi}{10},\sin\frac{\pi}{10})$ and
         $\hbar\omega=1.63J$ (lower eight panels) and
         at $(q^{x},q^{y})=6\frac{\pi}{a}(\cos\frac{\pi}{10},\sin\frac{\pi}{10})$ and
         $\hbar\omega=1.70J$ (upper eight panels)
         for Figure \ref{F:SqwLSWvsISW}(a${}^{\prime}$).
         ISW findings (the uppermost four panels and four panels in the third line)
         are compared with the $15$ types of subdomains $\tau(1)=1$ to $15$ ($\sqrt{5}Z=0,3$)
         or the $40$ types of subdomains $\tau(3)=1$ to $40$ ($\sqrt{5}Z=1,2$)
         illustrated with Figure \ref{F:Penrose}(d)
         (the four panels in the second line and lowermost four panels).}
\label{F:perpCP_ISWflats}
\end{figure}

   Figure \ref{F:perpCP_ISWflats} presents perpendicular-space mappings of 
$L\overline{S}(\bm{q};\omega)^{\![0]}\bigl|_{l}$ at
the same wavevector as Figure \ref{F:perpCP_LSWflat}
and energies $\hbar\omega=1.70J$ and $\hbar\omega=1.63J$
on the interaction-driven higher- and lower-lying split branches
[\ref{F:SqwLSWvsISW}(a${}^{\prime}$)] of the otherwise degenerate single nearly flat
scattering band.
Both spectral weightings still localize on the D and/or Q domains, evidencing that scattering
magnons to form these two separate nearly flat bands are still well confined within tricoordinated
sites, nay, more intriguingly claiming that they each are ``superconfined" within selected
tricoordinated sites.
The ISW ways of painting the D and Q domains are more complicated such that the Q domains are
divided into $\tau(1)$ subdomains, while the $\tau(1)=4$ subdomains in the D domains are further
partitioned into numerous $\tau(3)$ subdomains.
So far as the two ISW nearly flat modes are concerned, tricoordinated $\tau(0)=\mathrm{Q}$ sites
split into two types of subgroups $\tau(1)=5$ and $\tau(1)=6$, whereas tricoordinated $\tau(1)=4$
sites split into three types of subgroups $\tau(3)=10$ to $12$.
The spectral weights of the higher-lying nearly flat ISW scattering band center roughly on
the subdomain $\tau(1)=5$ with $\sum_{l'}z_{l'}(1)=16$,
while those of the lower-lying one center chiefly on the subdomains
$\tau(1)=6$, $\tau(3)=10$, and $\tau(3)=12$ all with $\sum_{l'}z_{l'}(1)=17$.
The $O(S^0)$ interactions prefer the higher-coordinated local environments labeled
$\tau(1)=4$ and $\tau(1)=6$ to the lower-coordinated ones labeled $\tau(1)=5$,
and they further choose the relatively high-coordinated local environments labeled
$\tau(3)=10$ and $\tau(3)=12$ among the $\tau(1)=4$ subdomains,
splitting the otherwise degenerate single nearly flat scattering band into two bands.
The lower (higher)-lying one is mediated by $O(S^0)$-interaction-stabilized (destabilized)
magnons ``superconfined" within particular tricoordinated sites surrounded by a larger (smaller)
number of bonds.
Confined antiferromagnons of $z=3$ character are robust against quantum fluctuations
and the $1/S$ corrections put them into ``superconfinement".

\section{Summary and Discussion}
\begin{figure}
\centering
\includegraphics[width=\linewidth]{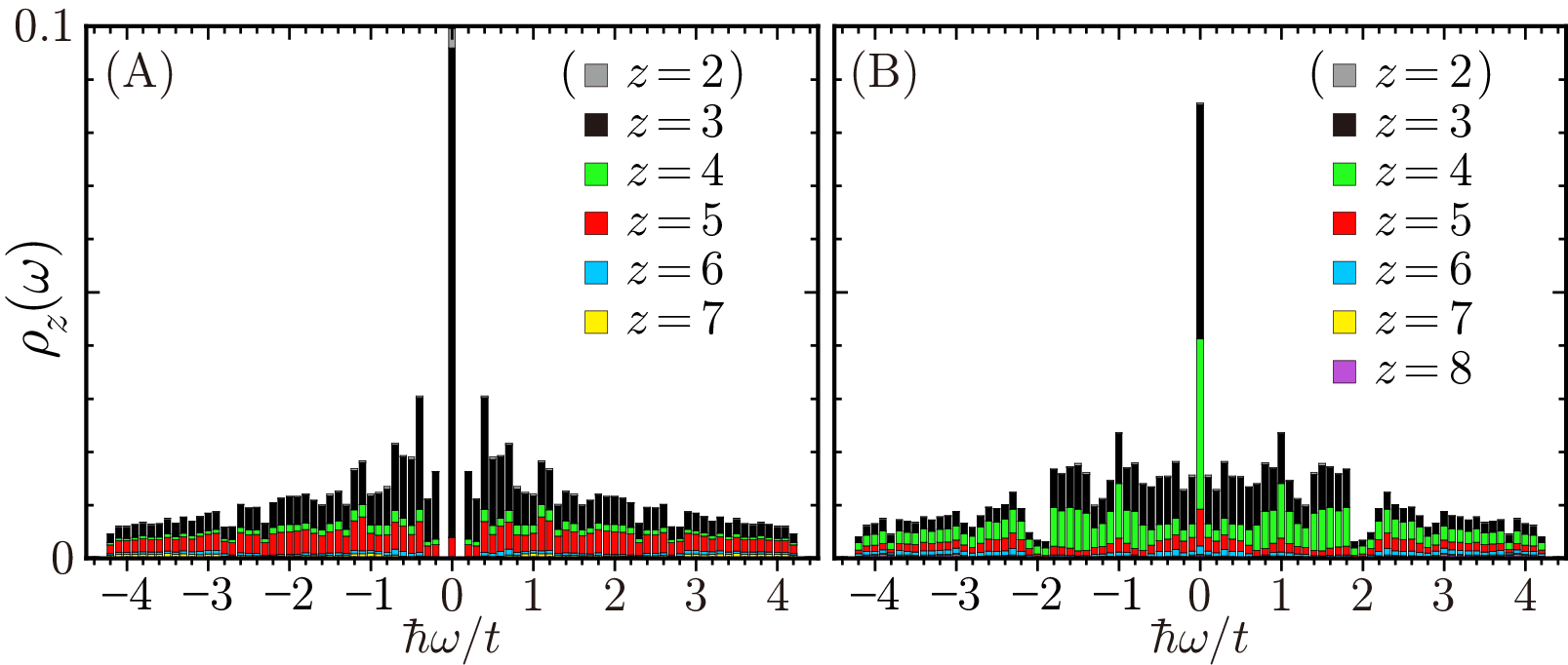}
\caption{%
         The site-resolved density of states $\sum_{z}\rho_{z}(\omega)$ of the fermionic excitation
         spectra, given by Eq. \eqref{E:TBDOS}, for the tight-binding Hamiltonian \eqref{E:TBHam}
         on the Penrose lattice of $L=11006$ (A) and Ammann-Beenker lattice of $L=10457$ (B).
         $z$ should take $3$ to $7$ and $3$ to $8$ on the Penrose and Ammann-Beenker lattices,
         respectively.
         Note that bicoordinated vertices are artifacts on finite clusters under the open
         boundary condition.}
\label{F:TBDOS}
\end{figure}

   Since the half-filled single-band Hubbard model reduces to our spin-$\frac{1}{2}$
antiferromagnetic Heisenberg model in question in its strong-correlation limit, it is worth while
to compare the above-revealed confined antiferromagnons with their fermionic origins.
A pioneering calculation \cite{K2740} was carried out for the vertex-model tight-binding
Hamiltonian
\begin{align}
   &
   \mathcal{H}_{t}
  =-t\sum_{\sigma=\uparrow,\downarrow}
     \sum_{i\in\mathrm{A}}\sum_{j\in\mathrm{B}}l_{i,j}
     (c_{i:\sigma}^{\dagger}c_{j:\sigma}+c_{j:\sigma}^{\dagger}c_{i:\sigma})
  =\sum_{\sigma=\uparrow,\downarrow}
   \mathbf{c}_{\sigma}^{\dagger}\mathbf{M}\mathbf{c}_{\sigma};
   \allowdisplaybreaks
   \nonumber \\
   &
   \bm{c}_{\sigma}^{\dagger}
  =\left[
    c_{1:\sigma}^{\dagger},\cdots,c_{L_{\mathrm{A}}:\sigma}^{\dagger},
    c_{L_{\mathrm{A}}+1:\sigma}^{\dagger},\cdots,c_{L_{\mathrm{A}}+L_{\mathrm{B}}:\sigma}^{\dagger}
   \right],\ 
   \allowdisplaybreaks
   \nonumber \\
   &
   \mathbf{M}
  =\left[
   \begin{array}{c|c}
     \mathbf{O}_{L_{\mathrm{A}}} & \mathbf{C}
    \\ \hline
     \mathbf{C}^{\dagger} & \mathbf{O}_{L_{\mathrm{B}}}
   \end{array}
   \right],\ 
   [\mathbf{O}_{L_{\mathrm{A}}}]_{i,i'}
  =[\mathbf{O}_{L_{\mathrm{B}}}]_{j,j'}
  =0,\ 
   [\mathbf{C}]_{i,j}=-tl_{i,j},
   \label{E:TBHam}
\end{align}
on the two-dimensional Penrose lattice,
where $c_{l:\sigma}^{\dagger}$ creates an electron with spin $\sigma$ at site $\bm{r}_{l}$
positioned on vertices of rhombuses,
$\bm{c}_{\sigma}^{\dagger}$ is a row vector of dimension $L$,
$\mathbf{O}_{L_{\mathrm{A}}}$ and $\mathbf{O}_{L_{\mathrm{B}}}$ are zero matrices
of dimension $L_{\mathrm{A}}\times L_{\mathrm{A}}$ and $L_{\mathrm{B}}\times L_{\mathrm{B}}$,
respectively.
An ordinary unitary transformation
\begin{align}
   \bm{c}_{\sigma}
  =\mathbf{U}\bm{\alpha}_{\sigma};\ \ 
   [\mathbf{U}]_{l,k}
  \equiv
   u_{l,k},\ 
   \bm{\alpha}_{\sigma}^{\dagger}
  \equiv
   \left[
     \alpha_{1:\sigma}^{\dagger},\cdots,\alpha_{L:\sigma}^{\dagger}
   \right]
   \label{E:unitarytransform}
\end{align}
rewrites \eqref{E:TBHam} into
\begin{align}
   \mathcal{H}_{t}
  =\sum_{\sigma=\uparrow,\downarrow}\sum_{k=1}^{L}
   \varepsilon_{k}
   \alpha_{k:\sigma}^{\dagger}\alpha_{k:\sigma},
   \label{E:TBdiagHam}
\end{align}
where $\alpha_{k:\sigma}^{\dagger}$ creates a quasiparticle fermion with spin $\sigma$
of energy $\varepsilon_{k}$.
While $\varepsilon_k$ is no longer a cosine band on any quasiperiodic lattice,
this Hamiltonian at half filling possesses the electron-hole symmetry---electrons with energy above
the Fermi sea behave the same as holes emergent in the Fermi sea.
It is nowadays well known that the Hamiltonian $\mathcal{H}_{t}$ yields macroscopically degenerate
confined states of $\varepsilon_k=0$ on not only the Penrose \cite{K2740,A1621,K214402} but also
Ammann-Beenker \cite{K115125,JR834} lattices.
In order to see which sites form these confined states, we plot in Figure \ref{F:TBDOS}
the site-resolved density of states
\begin{align}
   \rho(\omega)
  =\sum_{l=1}^{L}\rho_{l}(\omega)
  =\sum_{z=z_{\mathrm{min}}}^{z_{\mathrm{max}}}\rho_{z}(\omega)
  =\sum_{z=z_{\mathrm{min}}}^{z_{\mathrm{max}}}
   \frac{1}{L}\sum_{k=1}^{L}
   \frac{\sum_{l(z_l=z)}|u_{l,k}|^{2}}{\sum_{l=1}^{L}|u_{l,k}|^{2}}
   \delta\left(\hbar\omega-\varepsilon_{k}\right)
   \label{E:TBDOS}
\end{align}
of the quasiparticle fermion excitation spectrum.
For the Penrose lattice, the $\delta$-function peak at $\hbar\omega=0$ accounts for about
$10$ percent of the total number of states, consists of strictly localized---confined within
sites of $z=3$ and $z=5$---states, and separated from the remainder of the states---two symmetric
continuum---by a gap $0.172871t$ \cite{A1621,K214402}.
We find a macroscopically degenerate $\delta$-function-like peak at $\hbar\omega=0$ for
the Ammann-Beenker lattice as well.
While its constituent states involve all types of site, they each are still confined
\cite{K115125}.
All these similarities and differences between the Penrose and Ammann-Beenker lattices have
ever been observed in Figure \ref{F:HeisenbergDOS}.

\begin{figure}
\centering
\includegraphics[width=0.9\linewidth]{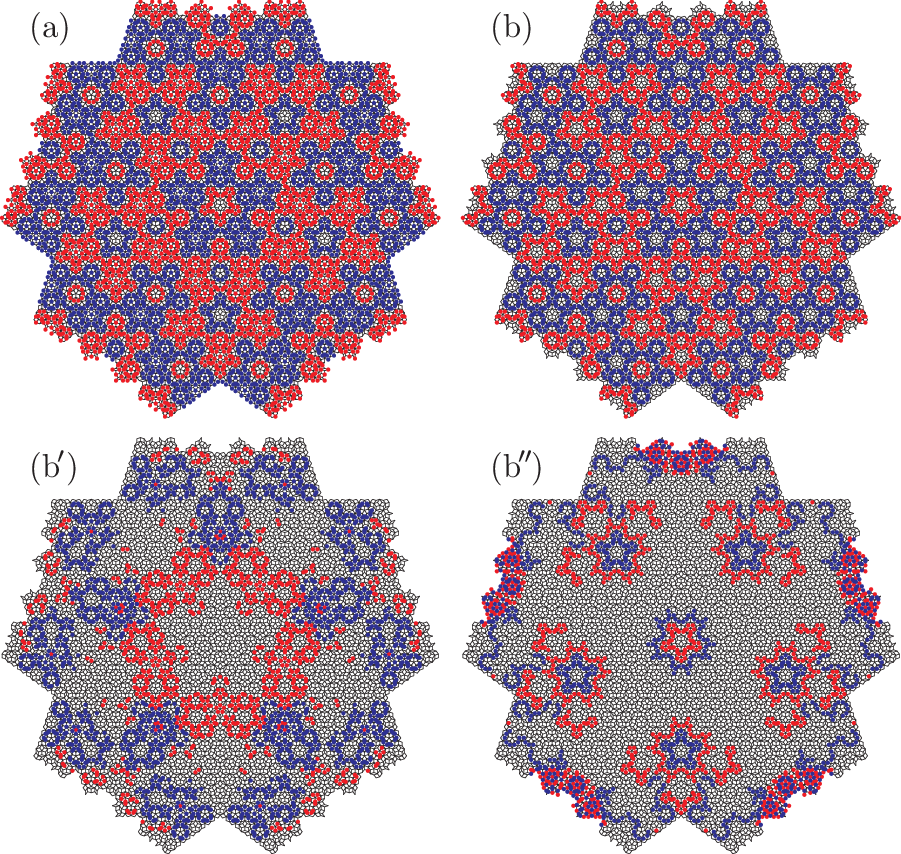}
\caption{%
         The site-resolved densities of states $\rho_{l}(\omega)$ \eqref{E:TBDOS} and
         $\rho_{l}(\omega)^{\![m]}$ \eqref{E:SWDOS} multiplied by the number of sites $L$ as
         functions of site position $\bm{r}_l$.
         The $l$th site is marked with a red ($l\in\mathrm{A}$) or blue ($l\in\mathrm{B}$) dot
         when $L\rho_{l}(\omega)$ is larger than or equal to $10^{-2}$.
         (a) $L\rho_{l}(\omega)$ at $\hbar\omega=0$
             for the tight-binding Hamiltonian \eqref{E:TBHam}.
             At every unmarked site $l$, $L\rho_{l}(\omega)<10^{-15}$.
         (b) $L\rho_{l}(\omega)^{\![1]}$ at $\hbar\omega=\frac{3}{2}J$
             for the LSW Hamiltonian \eqref{E:HBL[1]}.
             At every unmarked site $l$, $L\rho_{l}(\omega)^{\![1]}<10^{-15}$.
         (b${}^{\prime}$) 
             $L\rho_{l}(\omega)^{\![0]}$ summed up over the energies
             $1.63-10^{-3}\leq\hbar\omega/J\leq 1.63+10^{-3}$
             for the ISW Hamiltonian \eqref{E:HBL[0]}.
             At every unmarked site $l$, $L\rho_{l}(\omega)^{\![0]}<10^{-2}$.
         (b${}^{\prime\prime}$) 
             $L\rho_{l}(\omega)^{\![0]}$ summed up over the energies
             $1.70-10^{-3}\leq\hbar\omega/J\leq 1.70+10^{-3}$
             for the ISW Hamiltonian \eqref{E:HBL[0]}.}
\label{F:confined}
\end{figure}

   We can see at a glance the spatial confinement of the zero-energy degenerate states of
the tight-binding Hamiltonian \eqref{E:TBHam} on the Penrose lattice in Figure \ref{F:confined}(a).
All the confined states are separated from each other by
``forbidden ladders" \cite{A1621,K214402}---one-dimensional unbranched closed loops with
no wavefunction amplitude.
Every weighted domain enclosed by forbidden ladders reasonably consists either of tricoordinated
sites only or of pentacoordinated plus tricoordinated sites, that is in any case confined in either
of the two sublattices.
Every forbidden ladder lies between a couple of confined domains made of different sublattices.
Without any Coulomb interaction, these confined domains, i.e., tricoordinated and pentacoordinated
states, have staggered magnetizations.
On-site Coulomb repulsion $U$ emergent to the tight-binding Hamiltonian \eqref{E:TBHam}
immediately lifts the degeneracy among its zero-energy states and causes staggered magnetizations
on all sites of $z=3$ to $7$.
The local staggered magnetization on the site at $\bm{r}_l$ as a function of $U$,
$m_l(U)$, grows linearly with increasing $U$ and dependently on $l$,
$|m_l(U)-m_l(0)|\propto\kappa_l U$ \cite{K214402},
as long as $U$ is sufficiently small,
in contrast to the exponential growth free from site dependence with $U$ emergent to
a nonmagnetic tight-binding Hamiltonian on any bipartite lattice,
$|m_l(U)|\propto e^{-\kappa/U}$ \cite{F1999}.
Such an unconventional growth of staggered magnetizations can be seen on
the Ammann-Beenker \cite{K115125} and periodic Lieb \cite{N063622} lattices as well,
the latter of which exhibits a flat-band ferromagnetism,
$m_l(0)\neq 0$ only on bicoordinated sites of $z_l=2$.
On every quasiperiodic lattice, $\kappa_l$ depends on $l$ in a complicated manner because of
the lack of translational symmetry, but $\kappa_l$ increases with $z_l$ decreases.
Tricoordinated sites are magnetized much faster than any other site with increasing $U$
\cite{K214402,K115125}.
Thus and thus, the confined states made only of tricoordinated sites well survive the on-site
Coulomb repulsion.

   Here in the strong correlation limit, come the observations
Figures \ref{F:confined}(b), \ref{F:confined}(b${}^{\prime}$), and
\ref{F:confined}(b${}^{\prime\prime}$).
Longitudinal components in the exchange Hamiltonian \eqref{E:Ham} bring a local potential on
every site.
In the LSW Hamiltonian \eqref{E:ABCmatrices[1]}, it reads $z_{l}SJ$
[See Eq. \eqref{E:ABCmatrices[1]}]
to destabilize higher-coordinated sites more.
Interestingly enough,
Figure \ref{F:confined}(a) with all weighted pentacoordinated sites away coincides with
Figure \ref{F:confined}(b).
The alternate arrangement of A- and B-sublattice-confined domains with forbidden ladders
in between remains exactly the same between Figures \ref{F:confined}(a) and \ref{F:confined}(b).
Antiferromagnons are thus confined within only tricoordinated sites.
Bringing LSWs into interaction yields more intriguing observations,
Figures \ref{F:confined}(b${}^{\prime}$) and \ref{F:confined}(b${}^{\prime\prime}$),
where site potentials are no longer necessarily equal even among sites of the same coordination
number, because their surrounding environments become effective in the ISW Hamiltonian
\eqref{E:ABCmatrices[0]}.
ISWs yielding the lower- and higher-lying nearly flat scattering bands in
Figure \ref{F:SqwLSWvsISW} look complementary to each other in spatial distribution,
whose wavefunction amplitudes,
Figures \ref{F:confined}(b${}^{\prime}$) and \ref{F:confined}(b${}^{\prime\prime}$),
add up approximately to the LSW ones, \ref{F:confined}(b).
The lower- and higher-lying nearly flat scattering bands are mediated by antiferromagnons
stabilized and destabilized by $O(S^0)$ interactions, respectively, which are confined within
tricoordinated sites surrounded by larger and smaller numbers of bonds, respectively.
Quantum fluctuations encourage further confinement of antiferromagnons,
which may be referred to as ``superconfinement".
Note again that the nearly flat mode in the dynamic structure factor is characteristic of
the Penrose lattice and it splits in two due to the variety of its constituent local environments.
Superconfined antiferromagnons are thus peculiar to the two-dimensional Penrose lattice.

\vspace{6pt}







\appendix
\section*{Appendix:\ \  Dynamic Transverse and Longitudinal Structure Factors}
   Transverse and longitudinal contributions in the dynamic structure factors shown in
Figures \ref{F:SqwPenrose}(a), \ref{F:SqwPenrose}(a${}^{\prime}$),
\ref{F:SqwAB}(a), and \ref{F:SqwAB}(a${}^{\prime}$) are separately presented.
\begin{figure}[h]
\centering
\includegraphics[width=\linewidth]{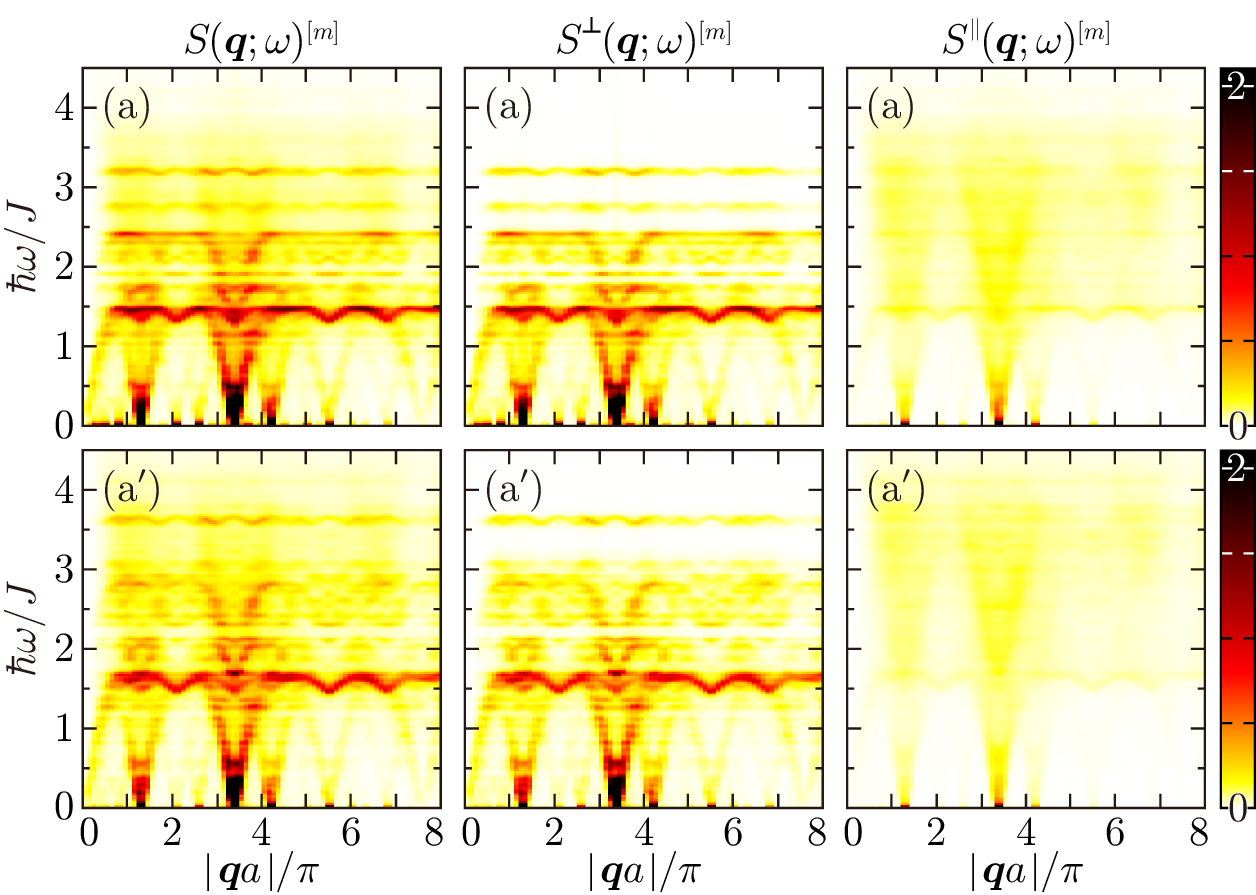}
\caption{%
         The dynamic structure factors
         $S(\bm{q};\omega)^{\![m]}
         \equiv
          S^\perp    (\bm{q};\omega)^{\![m]}
         +S^\parallel(\bm{q};\omega)^{\![m]}$
         in the LSW [$m=1$; (a)] and ISW [$m=0$; (a${}^{\prime}$)] ground states,
         given by Eqs. \eqref{E:SqwExpressionT} and \eqref{E:SqwExpressionL},
         for the Penrose lattice of $L=11006$
         along the path $(\mathrm{a})\equiv(\mathrm{a}^{\prime})$
         specified in Figure \ref{F:Sq}(A).
         Every spectral intensity is Lorentzian-broadened by a width of $0.01J$.}
\label{F:SqwTransLongiPenrose}
\end{figure}
\begin{figure}[h]
\centering
\includegraphics[width=\linewidth]{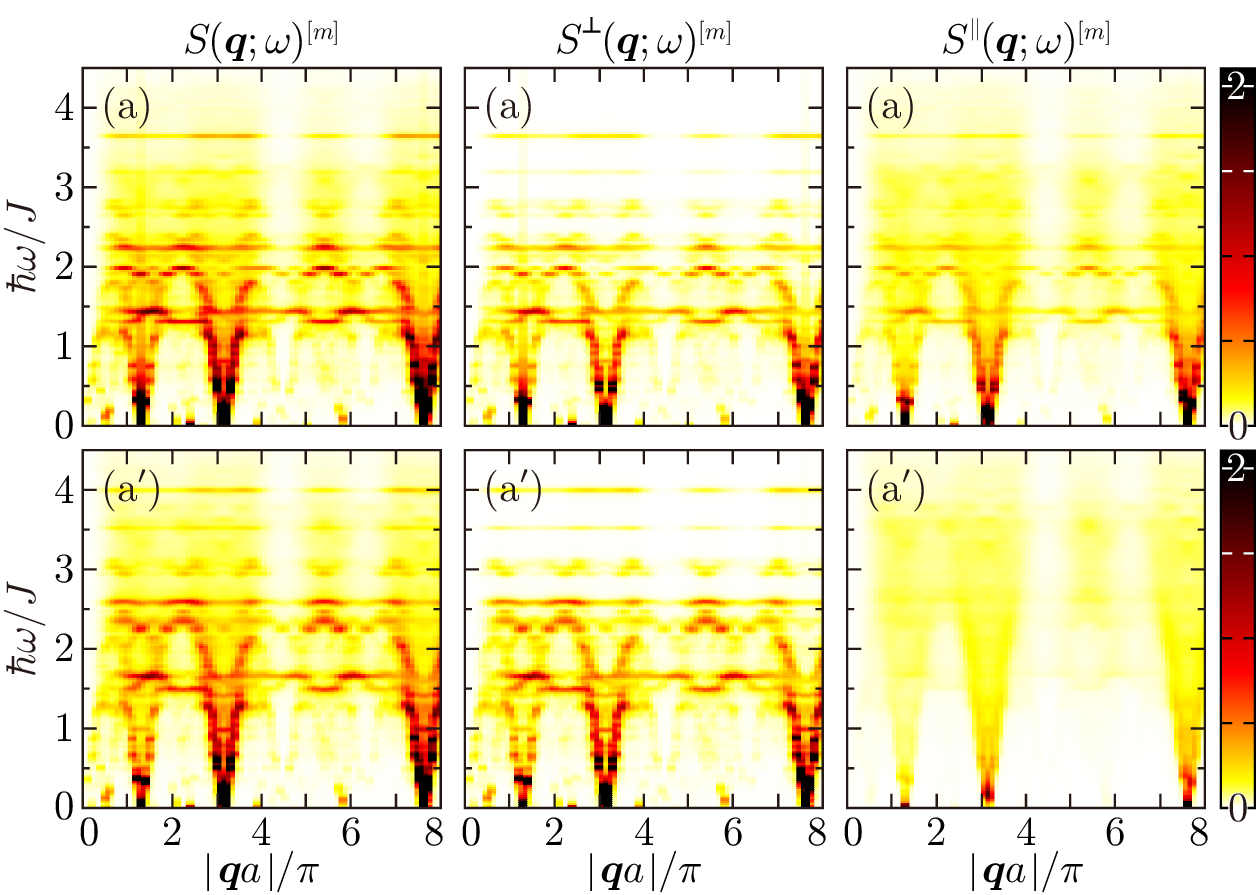}
\caption{%
         The dynamic structure factors
         $S(\bm{q};\omega)^{\![m]}
         \equiv
          S^\perp    (\bm{q};\omega)^{\![m]}
         +S^\parallel(\bm{q};\omega)^{\![m]}$
         in the LSW [$m=1$; (a)] and ISW [$m=0$; (a${}^{\prime}$)] ground states,
         given by Eqs. \eqref{E:SqwExpressionT} and \eqref{E:SqwExpressionL},
         for the Ammann-Beenker lattice of $L=10457$
         along the path $(\mathrm{a})\equiv(\mathrm{a}^{\prime})$
         specified in Figure \ref{F:Sq}(B).
         Every spectral intensity is Lorentzian-broadened by a width of $0.01J$.}
\label{F:SqwTransLongiAB}
\end{figure}

\end{document}